\PassOptionsToPackage{table}{xcolor}
\documentclass[]{aastex631}

\submitjournal{ApJ; Accepted October 21, 2025~$-$ \href{https://doi.org/10.3847/1538-4357/ae16a1}{doi:10.3847/1538-4357/ae16a1}}

\graphicspath{{./}{figures/}}
\usepackage[version=4]{mhchem}
\usepackage{multirow}
\usepackage{booktabs}
\usepackage{upgreek}
\usepackage{wasysym}
\usepackage[flushleft]{threeparttable}
\usepackage{comment} 
\usepackage{savesym}
\savesymbol{tablenum}
\usepackage{siunitx}
\restoresymbol{SIX}{tablenum}
\newcommand{\midsepremove}{\aboverulesep = 0mm \belowrulesep = 0mm}
\midsepremove
\newcommand{\midsepdefault}{\aboverulesep = 0.4ex \belowrulesep = 0.65ex}
\midsepdefault
\usepackage{array} 
\definecolor{red_bandpass}{RGB}{255, 0, 0}
\definecolor{bluered_bandpass}{RGB}{0, 0, 255}
\definecolor{vis_bandpass}{RGB}{255, 215, 0}
\definecolor{vNIR_bandpass}{RGB}{44, 160, 44}
\definecolor{NIR_bandpass}{RGB}{148, 103, 189}
\begin{document}

\title{Earth Analogs in Reflected Light: Insights from Early Spectral Characterization in Unconstrained Orbits}

\correspondingauthor{Arnaud Salvador}
\email{arnaud.salvador@dlr.de}

\author[0000-0001-8106-6164]{Arnaud Salvador}
\affiliation{Lunar \& Planetary Laboratory, University of Arizona, Tucson, AZ 85721, USA}
\affiliation{Department of Astronomy and Planetary Science, Northern Arizona University, Flagstaff, AZ 86011, USA}
\affiliation{Habitability, Atmospheres, and Biosignatures Laboratory, University of Arizona, Tucson, AZ 85721, USA}
\affiliation{NASA Nexus for Exoplanet System Science Virtual Planetary Laboratory, University of Washington, Box 351580, Seattle, WA 98195, USA}
\affiliation{Institute of Space Research, German Aerospace Center (DLR), Rutherfordstr. 2, 12489 Berlin, Germany}

\author[0000-0002-3196-414X]{Tyler D. Robinson}
\affiliation{Lunar \& Planetary Laboratory, University of Arizona, Tucson, AZ 85721, USA}
\affiliation{Department of Astronomy and Planetary Science, Northern Arizona University, Flagstaff, AZ 86011, USA}
\affiliation{Habitability, Atmospheres, and Biosignatures Laboratory, University of Arizona, Tucson, AZ 85721, USA}
\affiliation{NASA Nexus for Exoplanet System Science Virtual Planetary Laboratory, University of Washington, Box 351580, Seattle, WA 98195, USA}

\begin{abstract}

A next generation of space-based observatories aims to detect and characterize potentially Earth-like exoplanets around Sun-like stars using reflected light spectroscopy.
However, it remains unclear how such direct imaging observations---limited in spectral coverage and signal-to-noise ratio (S/N)---translate into constraints on atmospheric composition and habitability.
Coronagraphs used for high-contrast imaging typically operate over narrow bandpasses, and exposure time limits can restrict data quality.
To optimize observing strategies and instrument design, we use our atmospheric retrieval tool, \texttt{rfast}, to assess the performance of a \textit{Habitable Worlds Observatory}-type mission across different spectral bandpasses (``Red'', ``Blue'', ``Visible'', ``NIR'', and their combination) and S/N levels (10, 15, and 20; from moderate to moderate-high observation quality) in retrieving a wide range of 17 atmospheric, surface, bulk, and orbital parameters of a habitable Earth analog. We outline the observation requirements for each parameter and the detection capabilities of each case, within a novel scenario where spectral data are taken ``early'', prior to achieving orbit constraints (which may require repeat visits to a system).
For coronagraph-restricted and NIR-only bandpasses, most of the limited retrievable information is already captured at S/N = 10, with little improvement at higher S/N.
For broader spectral coverage, the quality and quantity of retrieved information improve with increasing S/N, but combining visible and NIR ranges provides the most comprehensive characterization, even at moderate S/N.
To maximize returns, wider spectral coverage should be prioritized over improving S/N when spectral access is limited.

\end{abstract}

 
\section{Introduction} \label{sec:intro}

With over 5900 confirmed exoplanets\footnote{\url{https://exoplanetarchive.ipac.caltech.edu}}---more than a third of which are rocky---detecting and characterizing habitable worlds \citep[i.e., where liquid water is stable at the planetary surface;][]{Kasting1993b, Kopparapu2013} has become a major goal for the next decade \citep{Decadal2021, Decadal_strategy2022}.
Direct imaging and reflected light spectroscopy are promising methods for detailed characterization of exoplanetary atmospheres and surface conditions, which are critical for assessing the potential habitability of a world.
However, such observations require starlight suppression techniques (e.g., coronagraphs or starshades) to achieve the necessary $<10^{-10}$ planet-to-star flux contrast for direct imaging of Earth-like planets around Sun-like stars \citep[orbiting at 1 au and observed in optical wavelengths; e.g.,][]{Guyon2006, Seager2010}.
NASA's upcoming \textit{Nancy Grace Roman Space Telescope}\footnote{\url{https://roman.gsfc.nasa.gov}} will be a major milestone in space-based coronagraphy, demonstrating high-contrast imaging capabilities with unprecedented observations of nearby giant exoplanets \citep{Bailey2018, Akeson2019, Mennesson2020, Kasdin2020, Mennesson2022}. This mission will pave the way for the direct imaging of exo-Earths with NASA's proposed \textit{Habitable Worlds Observatory} (HWO; building upon the earlier HabEx and LUVOIR mission concepts, e.g., \citealp{Gaudi2018, Habex2020, Roberge2018, Luvoir2019}).

To design exo-Earth direct imaging missions and assess the returns of proposed instruments and observing scenarios, atmospheric retrieval models are commonly used \citep[e.g.,][]{Madhusudhan2018, Fortney2021_chapter}. These models infer the range of exoplanet properties that can explain either actual or synthetic observations and can thus assess how future instruments and observations will translate into constraints on these parameters.

Recent studies have focused on the characterization of gas and ice giants in reflected light including with the upcoming Roman Space Telescope \citep[e.g.,][]{Marley2014, Lupu2016, Nayak2017, Carrion-Gonzalez2020, Damiano2020a, Damiano2020, Carrion-Gonzalez2021}, as well as on Earth-twins imaging \citep[e.g.,][]{Feng2018, Smith2020, Damiano2022, Robinson2023, Hall2023, Susemiehl2023, Kofman2024}. These efforts typically target the detectability of specific parameters such as habitability indicators and biosignature gases \citep[e.g.,][]{Wang2018a, Latouf2023a, Latouf2023b, Latouf2025, Metz2024, Tokadjian2024, Currie2025} or the inference capabilities under specific observing scenarios \citep{Damiano2023, Young2024a, Salvador2024, Damiano2025}.
Parallel studies have explored habitable planet characterization using transmission spectroscopy \citep[e.g.,][]{Benneke2012, Greene2016, Min2020, Pidhorodetska2020, Gialluca2021, Currie2023, Meadows2023, Lustig-Yaeger2023}, thermal emission \citep[e.g.,][]{VonParis2013b, Stevenson2020, Lustig-Yaeger2021, Fujii2021, Mandell2022, Konrad2022, Alei2022, Mettler2024}, and potential synergies across different techniques and missions \citep[e.g.,][]{Gilbert-Janizek2024, Alei2024}.

However, a persistent challenge in atmospheric retrievals is the degeneracy between inferred parameters \citep[e.g.,][]{Welbanks2019, Barstow2020, Novais2025}: multiple combinations of parameters (e.g., cloud properties, planetary radius, surface albedo, phase angle, orbital distance) can produce similar spectra, making it difficult to disentangle their individual influences \citep[e.g.,][]{Nayak2017, Wang2022_surface, Salvador2024}.
Constraining certain parameters can help break degeneracies and enable/enhance the inference of correlated ones \citep[e.g., constraining the orbital parameters allows to infer the planetary radius, which itself refines clouds optical depth and methane inferences;][]{Nayak2017, Carrion-Gonzalez2020, Carrion-Gonzalez2021, Salvador2024}, but risks introducing artificial bias when only a limited number of parameters are inferred.
A fully agnostic approach, where all relevant parameters remain unconstrained unless informed by existing data, is also essential for robust target identification during an initial blind-search phase before follow-up observations \citep[e.g.,][]{Luvoir2019}, in a self-contained survey \citep[e.g.,][]{Habex2020}, or to assess the full capabilities of HWO in its different observing phases. This approach ensures that the analysis reflects the uncertainty inherent in exoplanet observations, where most parameters are unknown and need to be explored.
To better reflect the complexity of actual observations and minimize potential biases, inverse modeling should not focus solely on habitability indicators but rather aim to retrieve as many spectrally relevant parameters as possible.

Furthermore, modeling efforts must contend with the generally detrimental effects of clouds and hazes. While hazes can possibly inform planetary composition and processes \citep[e.g.,][]{Corrales2023}, and clouds improve the detectability of atmospheric constituents in specific cases \citep[e.g.,][]{Kelkar2025}, they can mute absorption features of key atmospheric species \citep[e.g.,][]{Fauchez2019, Komacek2020, Pidhorodetska2020, Barstow2020, Kelkar2025}. Broader wavelength coverage can mitigate the resulting degeneracies between gas abundances and cloud properties \citep[e.g.,][]{Fauchez2019, Barstow2020}, but observing time limitations and instrumental constraints---such as the restricted bandpass of coronagraphs---reduce achievable S/N and spectral coverage. These observations may miss key spectral information, leading to biased retrievals of atmospheric composition and cloud properties \citep[e.g.,][]{Mai2019, Kelkar2025}.
In addition, if the latter are not retrieved, fixed cloud parameterizations may become inconsistent with sampled atmospheric states, failing to capture their spectral effects and leading to unrealistic results. For example, if water vapor abundance is retrieved without accounting for corresponding changes in water cloud parameterization \citep[as considered in][which ensures physical consistency between water vapor and cloud formation]{Damiano2023, Damiano2025}, inconsistencies between the sampled values of water vapor and the spectral imprints of water clouds could bias the retrievals and compromise their validity. While \citet{Barstow2020b} demonstrated its moderate influence on the retrieved abundance of \ce{H2O} from transmission spectroscopy, this effect has not been investigated for reflected light observations and a broader range of retrieved parameters, which should further be addressed in a dedicated study.

Focusing on isolated parameters or narrow bandpasses can therefore bias retrieval analysis and fail to capture the broader challenges of exoplanet characterization under limited observational knowledge.
Despite recent advances, it remains unclear how direct imaging observations---limited by spectral coverage and observational noise---translate into constraints on atmospheric composition and key indicators of habitability. A systematic comparison of retrieval performance across different spectral ranges, S/N levels, and a large number of unconstrained parameters is still lacking. 
A critical challenge is defining the most efficient observing strategies to robustly identify Earth analogs. One possible approach involves an initial large-scale survey using time-limited, lower S/N observations in a reduced bandpass focused on water vapor detection, followed by a more detailed characterization of the most promising candidates with full spectral capabilities \citep[e.g.,][]{Luvoir2019}.
In this scenario, it is essential to understand (1) what constraints can be obtained from reduced-bandpass, low (to high) S/N observations when most parameters are unknown, and (2) how these constraints inform or bias spectral interpretation.
This work aims to fill this gap by systematically evaluating how spectral coverage (including coronagraph-restricted bandpasses) and S/N impact the identification of Earth-like habitable worlds. Specifically, we investigate how these factors influence the retrieval of atmospheric and bulk properties of an Earth analog observed in reflected light without prior determination of the orbit.
In doing so, we aim to provide key insights that will help inform the design of future direct imaging missions, such as HWO, and optimize observing strategies to balance spectral completeness, observing time, and scientific yield.

Section~\ref{sec:methods} describes our retrieval tool \texttt{rfast}---designed to explore high-dimensional parameter spaces efficiently without prohibitive computational costs \citep[e.g.,][]{Robinson2023, Hall2023, Salvador2024}---as well as the inferred parameters and the observing scenarios considered. Section~\ref{sec:results} presents our findings, followed by a discussion of their implications for future exo-Earth characterization with HWO in Section~\ref{sec:discuss}. Finally, key conclusions are summarized in Section~\ref{sec:conclusion}.

\section{Methods} \label{sec:methods}
In this section, we provide an overview of the \texttt{rfast} atmospheric retrieval tool and the setup used to derive key properties from simulated reflected-light observations of Earth analogs. We also describe the observation scenarios, including the selection of spectral bandpasses and signal-to-noise ratio (S/N) considered for this study.

\subsection{The \texttt{rfast} Atmospheric Retrieval Tool and Setup}
We use \texttt{rfast} \citep{Robinson2023, Salvador2024}, a versatile atmospheric retrieval framework, to assess inferences of 17 unknown/free atmospheric and planetary properties (\autoref{tab:retrieved_pars}) out of synthetic observations of an Earth-analog orbiting a Sun-like star through various spectral bandpasses and S/N values (\autoref{tab:retrieval_grid}) proposed for HWO.

Future HWO observations will collect the light from the host star reflected by the planet directly imaged as a resolved and distinct light source. At a given star-planet-observer phase angle, $\alpha$, it will measure the wavelength-dependent planet-to-star flux ratio, $F_{\rm p}/F_{\rm s}$, defined as:
\begin{equation}
    \frac{F_{\rm p}}{F_{\rm s}} = A_{\rm g} \Phi (\alpha) \left( \frac{R_{\rm p}}{a} \right)^{2} \ ,
    \label{eq:Fp_Fs}
\end{equation}
where $A_{\rm g}$ is the geometric albedo, $\Phi$ is the phase function (which depends on the phase angle), $R_{\rm p}$ is the radius of the planet, and $a$ is the orbital distance.
The phase angle dependence, which defines the viewing angle geometry, is accounted for in \texttt{rfast} 3D mode \citep{Robinson2023}. The planet is treated as a homogeneous pixelated globe (such as a disco ball), where radiative transfer calculations are performed for each pixel under local plane-parallel approximation. Depending on the illumination geometry, each facet of the pixelated globe has different pairs of incidence and emergence angles. A Gauss-Chebyshev integration of the flux emitted by all plane-parallel facets gives the total emergent flux from the spatially-integrated illuminated disk/planet \citep[e.g.,][]{Horak1965, Robinson2023, Salvador2024}.

\begin{figure}
    \centering
    \includegraphics[width=.5\textwidth, height=1.\textheight, keepaspectratio]{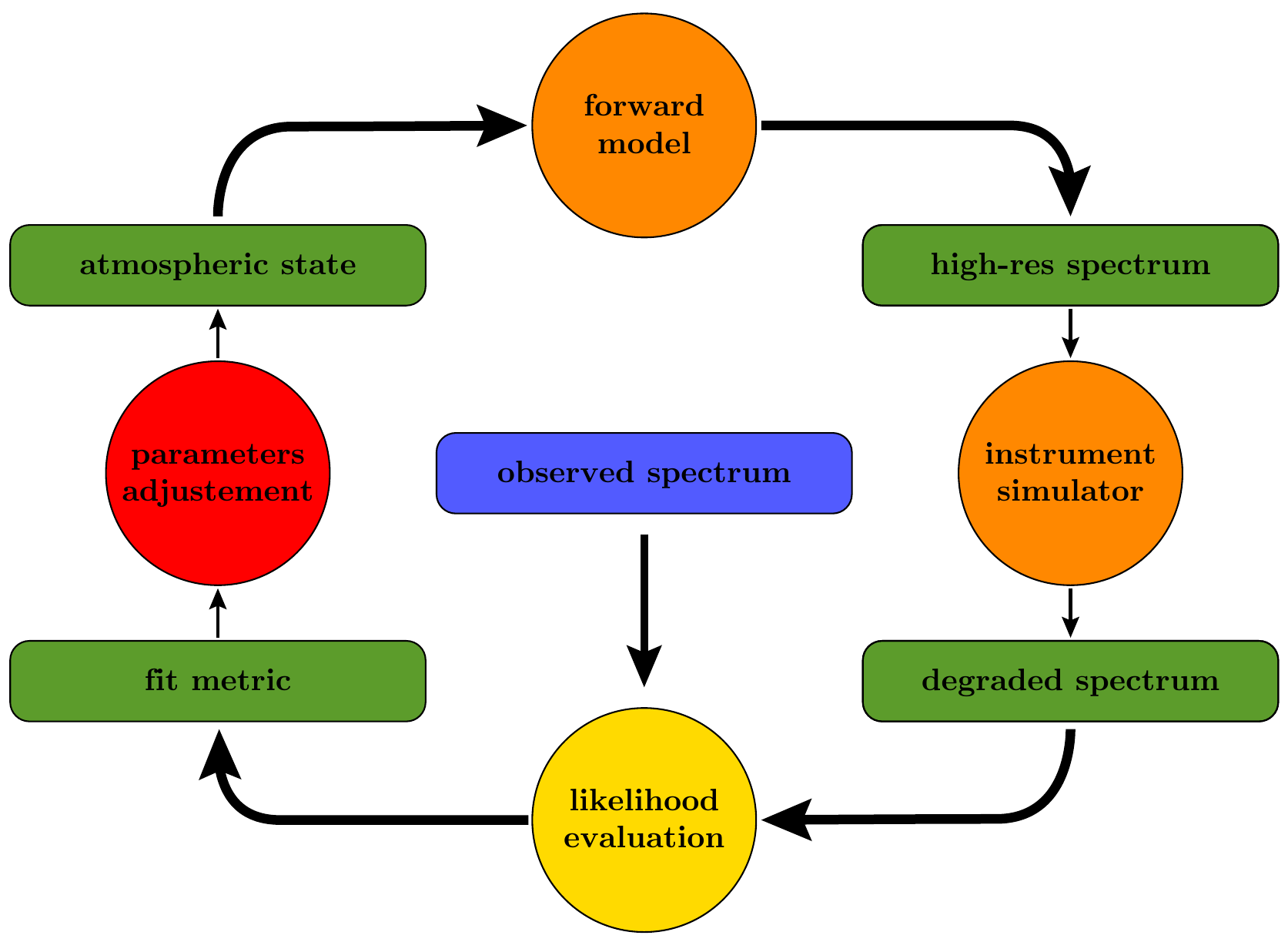}
    \caption{Schematic of the \texttt{rfast} retrieval framework. Boxes denote model inputs and outputs, while circles represent components of the framework. Here, the ``observed'' spectrum is a synthetic observation of an Earth analog, generated using Earth-based fiducial parameters (see \autoref{tab:retrieved_pars}).
    }
    \label{fig:retrieval_algorithm}
\end{figure}

Since the overriding goal of HWO is the characterization of Earth-like habitable exoplanets, we focus on observations of the best characterized habitable planet: Earth.
The retrieval algorithm is depicted in \autoref{fig:retrieval_algorithm} and proceeds as follows. A radiative transfer ``forward'' model first generates a high-resolution reflected light spectrum of Earth based on its known (fiducial) properties (\autoref{tab:retrieved_pars}), i.e., with an atmospheric composition dominated by \ce{N2} (78\%) and \ce{O2} (21\%), with \ce{H2O}, \ce{CO2}, \ce{CH4}, and \ce{O3} as additional trace species, and \ce{Ar} as the background gas. Note that the forward model-generated reflected light spectrum has been validated against Earth observations \citep{Robinson2023}.
An instrument model then degrades the resolution and adds noise to mimic a ``faux observation'' (\autoref{fig:ref_spectrum}) for a given HWO design and for a given observation scenario (\autoref{subsec:observation_scenarios}). These synthetic observations are then compared to spectra generated in the same way, but using sets of free parameters (consistent with any specified prior ranges) sampled by a Bayesian tool (\texttt{emcee}; \citealp{Foreman-Mackey2013}). The likelihood of the sampled parameter values is assessed by how well the newly generated spectra reproduce the faux observations (\autoref{fig:ref_spectrum}), via a standard chi-squared ($\chi^2$). When combined with the priors, this likelihood yields a posterior probability distribution for the parameters, which then informs how parameter space should be further explored.
After a burn-in period, the posterior probability distribution of each free parameter indicates the parameter ranges that can explain the observations.
To estimate remote sensing capabilities and accuracy of different observing scenarios, we repeat the same procedure considering observations for different spectral bandpasses and S/N (\autoref{tab:retrieval_grid}), and compare the retrieval results to one another.

Similar to \citet{Salvador2024}, the 17 retrieved parameters (whose prior ranges and Earth-based fiducial values are listed in \autoref{tab:retrieved_pars}) define the surface conditions via the surface pressure ($p_{\rm surf}$) and albedo ($A_{\rm surf}$), the atmospheric thermal and chemical state via a characteristic atmospheric temperature ($T$; which is probed through the temperature dependence of the band shapes) and the atmospheric volume mixing ratios of \ce{N2}, \ce{O2}, \ce{H2O}, \ce{CO2}, \ce{CH4}, and \ce{O3} (with \ce{Ar} back-filling the rest of the atmosphere when the total volume mixing ratio is lower than one), as well as the cloud, orbital, and planetary bulk properties. The cloud properties are defined via the cloudiness fraction ($f_{\rm c}$) and their top pressure ($p_{\rm c}$), thickness ($\Delta p_{\rm c}$), and optical depth ($\tau_{\rm c}$). The planetary bulk properties are the planet mass ($M_{\rm p}$) and radius ($R_{\rm p}$), and the orbital distance ($a$) and the planetary phase angle ($\alpha$) are the orbital parameters.

The retrieval settings were as follows. The \texttt{rfast} Bayesian sampling tool \texttt{emcee} was set with 15 Markov Chain Monte Carlo chains (walkers) per parameter (for a total of 255 walkers). Starting from random values around the truth, we let \texttt{emcee} explore the parameter space within the range of values defined as flat priors (\autoref{tab:retrieved_pars}) for 100k steps \citep[proven to be enough for the chains to fully converge for all retrieved parameters;][]{Robinson2023, Salvador2024}. We then draw the parameter posterior distributions from the last 5k steps, which gives a statistically representative sample without being computationally prohibitive. This is consistent with our previous studies \citep{Feng2018, Robinson2023, Salvador2024} and ensures thorough exploration of the parameter space by the walkers, regardless of their initial position.

\subsection{Observation Scenarios: Spectral Bandpasses Selection and S/N}\label{subsec:observation_scenarios}
Our observation scenarios consist of combinations of various spectral bandpasses and qualities proposed for typical HWO/LUVOIR/HabEx setups (\autoref{tab:retrieval_grid}). We consider five different observation scenarios---``Red'', ``Blue \& Red'', ``Visible'', ``vNIR'', and ``NIR'' (see \autoref{tab:retrieval_grid})---spanning both broad and narrow wavelength coverages to match instrument limitations and/or observation strategies that restrict or enhance access to spectral features indicative of habitability \citep[e.g.,][]{Luvoir2019, Young2024a}.
Aligning with technical restrictions of coronagraphs, which operate over relatively limited bandpasses (e.g., $\rm \lesssim 0.2~\upmu m$-wide ranges in the visible), we include scenarios with either a single 20\% coronagraph ``Red'' bandpass at red wavelengths ($\rm \lambda = [0.87 - 1.05]~\upmu m$, spectral resolving power $res = \lambda/\delta\lambda = 140$; targeting the prominent 0.94$~\upmu \rm m$ water vapor band) or two 20\% coronagraph ``Blue \& Red'' bandpasses---one at red wavelengths and one at blue wavelengths ($\rm \lambda = [0.43 - 0.53]~\upmu m$, $res = 140$; highlighting Rayleigh scattering slopes, which help constrain planet size).
A single red channel aligns with an early step in the LUVOIR ``search for life'' strategy focused on detecting water vapor signatures \citep[see, e.g., the 0.89 $\rm\upmu m$ bandpass in][]{Young2024a}, while the pair of Blue \& Red channels corresponds to later steps emphasizing atmospheric characterization \citep[see Figure 1.5 in][]{Luvoir2019}. Note that, unless they are in parallel coronagraphs, these single channels must be observed sequentially \citep[e.g.,][]{Luvoir2019, Young2024a}.
Beyond these restricted channels, we consider a ``Visible'' case with a broad bandpass (achievable with parallel coronagraph instruments or a starshade; \citealp{Shaklan2023, Shaklan2024, Mennesson2024}) covering the full visible/optical range ($\rm \lambda = [0.45 - 1.0]~\upmu m$, $res = 140$) for spectroscopic characterization and minimizing false negative detections of life. An additional optimistic, very broad ``vNIR'' bandpass is included, spanning visible to near-infrared wavelengths ($\rm \lambda = [0.45 - 1.8]~\upmu m$, with $res = 70$ in the NIR), consistent with in-depth follow up characterization of promising targets \citep{Luvoir2019}. Such wide spectral coverage could be acquired either simultaneously via parallel coronagraph instruments for maximum efficiency \citep[or sequentially for maximum sensitivity;][]{Luvoir2019}, or via a single radial movement of the starshade to access near-infrared wavelengths in a HabEx-type architecture \citep{Habex2020}.
Lastly, a ``NIR''-restricted scenario (spanning $\rm \lambda = [1.0 - 1.8]~\upmu m$, $res = 70$) is also considered to assess the value of these wavelengths.

\begin{figure}
    \centering
    \includegraphics[width=.5\textwidth, height=1.\textheight, keepaspectratio]{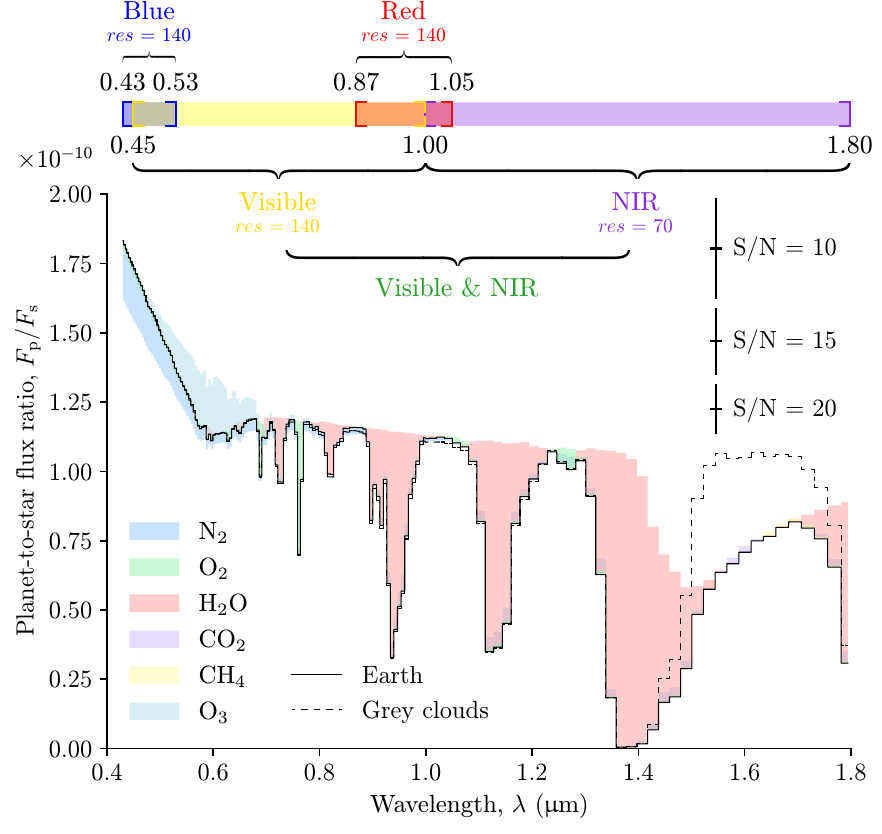}
    \caption{Model-generated (synthetic) reflected light spectrum of our fiducial case (plain line): an Earth-like planet at quadrature (i.e., at a planetary phase angle $\alpha=90$\textdegree; see \autoref{tab:retrieved_pars} for fiducial input values), and assuming grey cloud properties (dashed line), which notably do not capture the absorption feature of water ice clouds near 1.6 $\upmu$m. The colored areas indicate the spectral imprints of the gas species by showing the difference between the baseline spectrum (plain line) and the flux measured if they were absent from the atmosphere and replaced with Ar, the background gas.
    The top of the figure indicates the spectral ranges considered, along with their resolving power $res = \lambda/\delta\lambda$ (\autoref{tab:retrieval_grid}): Blue ($\rm \lambda = [0.43 - 0.53]~\upmu m$, $res = 140$) \& Red ($\rm \lambda = [0.87 - 1.05]~\upmu m$, $res = 140$) bandpasses, Visible ($\rm \lambda = [0.45 - 1]~\upmu m$, $res = 140$) \& near-infrared (NIR; $\rm \lambda = [1 - 1.80]~\upmu m$, $res = 70$) bandpasses (together denoted vNIR), or Red, Visible, and NIR bandpasses observed individually.
    The error bars represent the nonrandomized, wavelength-independent noise levels corresponding to specific signal-to-noise ratios (S/N), set at the lower bound of each spectral coverage (see \autoref{sec:appendix_SNR} and \autoref{fig:spectrum_SNR}): $\rm \lambda_0 = 0.43~\upmu m$ for the Blue \& Red bandpass, $\rm \lambda_0 = 0.45~\upmu m$ for the Visible and Visible \& NIR spectral ranges, $\rm \lambda_0 = 0.87~\upmu m$ for the Red bandpass, and $\rm \lambda_0 = 1.00~\upmu m$ for the NIR bandpass.
    }
    \label{fig:ref_spectrum}
\end{figure}

For each observation scenario, we consider three different spectral qualities (i.e., S/N, that translates into exposure times for a specific instrument design), spanning the range from moderate (S/N = 10) to moderate-high quality (S/N = 20), with an intermediate case (S/N = 15).
As in \citet{Salvador2024} and following \citet{Feng2018} initial validation, the noise is considered constant, nonrandomized (i.e., the error bars are simply centered on truth/noise-free values) and gray (i.e., wavelength-independent) to account for different data quality without being attached to specific observing parameters such as telescope diameter or target distance \citep[as opposed to integration times; see][]{Feng2018}.
The S/N value (10, 15, or 20) is set at the lower bound of each spectral bandpass (at $\lambda_{0} =$ 0.87, 0.43, and 1.0 $\rm ~\upmu m$ for the Red, Blue \& Red, and NIR scenarios, respectively, and at $\lambda_{0} = 0.45~\upmu \rm m$ for both the Visible and vNIR spectral bandpasses), and propagated across the entire spectral coverage while maintaining a constant noise/error bar size (shown on \autoref{fig:ref_spectrum} for each S/N value; see \autoref{sec:appendix_SNR} for more details). The corresponding S/N value obtained as a function of wavelength for each spectral coverage is shown in \autoref{fig:spectrum_SNR}.
On top of enabling easier reproducibility because of the wavelength independence, a constant gray noise aligns with the predicted performance of the HabEx and LUVOIR missions, assuming equal exposure times across their UV, optical, and NIR spectral bands \citep{Luvoir2019, Habex2020}. Furthermore, \citet{Feng2018} demonstrated that for a statistical sampling of retrievals, a more realistic, randomized noise instance would yield similar inference results.

A novel approach adopted for the retrievals presented here assumes that spectral observations are acquired ``early,'' meaning before orbit determination. In the absence of prior orbit constraints from precision radial velocity or host stellar astrometry, typical mission strategies involve multiple repeat photometric high-contrast imaging visits to planetary systems to constrain orbital parameters and identify worlds within the habitable zone of their host star. The ``early'' spectral characterization approach explored here represents a paradigm where spectroscopy is performed immediately after photometric detection, trading orbital information for the potential of earlier atmospheric species detection. One might call this the ``spectrum first, ask questions sooner'' approach.

With five variations of spectral coverages and three different S/N (\autoref{tab:retrieval_grid}), we considered a total of 15 different observation scenarios for which we ran retrieval analysis to assess their respective value in inferring the properties of Earth analogs and thus define the best observing strategy. The retrieval results are presented in the next section.

\begin{table}
\centering
\caption{Retrieved Parameters, Corresponding Description, Earth-based Fiducial Input Value, and Prior Range}
\label{tab:retrieved_pars}
\resizebox{0.5\textwidth}{!}{%
\begin{tabular}{@{}lllc@{}}
\toprule 
Parameter              & Description                         & Input                 & Flat Prior     \\
\midrule 
\multicolumn{4}{c}{\textit{Surface Conditions}}                                                       \\
log $p_{\rm surf}$     & Surface pressure (Pa)               & log($10^5$)           & [0,~8]         \\
$T$                    & Atmospheric temperature$^*$ (K)     & 255                   & [100,~1000]    \\
log $A_{\rm surf}$     & Surface albedo                      & log(0.05)             & [$-2$,~0]      \\
\midrule 
\multicolumn{4}{c}{\textit{Gas Abundances$^\dag$}}                                                    \\
log $f_{\rm\ce{N2}}$   & Molecular nitrogen mixing ratio     & log(0.78)             & [$-10$,~0]     \\
log $f_{\rm\ce{O2}}$   & Molecular oxygen mixing ratio       & log(0.21)             & [$-10$,~0]     \\
log $f_{\rm\ce{H2O}}$  & Water vapor mixing ratio            & log($3\times10^{-3}$) & [$-10$,~0]     \\
log $f_{\rm\ce{CO2}}$  & Carbon dioxide mixing ratio         & log($4\times10^{-4}$) & [$-10$,~0]     \\
log $f_{\rm\ce{CH4}}$  & Methane mixing ration               & log($2\times10^{-6}$) & [$-10$,~0]     \\
log $f_{\rm\ce{O3}}$   & Ozone mixing ratio                  & log($7\times10^{-7}$) & [$-10$,~$-2$]  \\
\midrule 
\multicolumn{4}{c}{\textit{Planetary Bulk Parameters}}                                                \\
log $R_{\rm p}$        & Planet radius ($R_{\oplus}$)        & log(1)                & [$-1$,~$1$]    \\
log $M_{\rm p}$        & Planet mass ($M_{\oplus}$)          & log(1)                & [$-1$,~$2$]    \\
\midrule 
\multicolumn{4}{c}{\textit{Cloud Parameters}}                                                         \\
log $p_{\rm c}$        & Cloud-top pressure (Pa)             & log($6\times10^{4}$)  & [0,~8]         \\
log $\Delta p_{\rm c}$ & Cloud thickness (Pa)                & log($10^4$)           & [0,~8]         \\
log $\tau_{\rm c}$     & Cloud optical depth                 & log(10)               & [$-3$,~$3$]    \\
log $f_{\rm c}$        & Cloudiness fraction                 & log(0.5)              & [$-3$,~0]      \\
\midrule 
\multicolumn{4}{c}{\textit{Orbital Parameters}}                                                       \\
$a$                    & Planetary orbital distance (AU)     & 1                     & [0.1,~10]      \\
$\alpha$               & Planetary phase angle (\textdegree) & 90                    & [0,~180]       \\
\bottomrule 
\multicolumn{4}{l}{$^*$Isothermal atmosphere temperature.}\\
\multicolumn{4}{l}{$^\dag$The remaining atmosphere is back-filled with argon.}\\
\end{tabular}%
}
\end{table}

\begin{table}
\caption{Retrieval Grid: Wavelength Coverages and Spectral Qualities Explored in This Study
}
\label{tab:retrieval_grid}
\centering
\resizebox{0.6\textwidth}{!}{\begin{tabular}{l c c}
\toprule 
\multicolumn{3}{c}{\textit{Wavelength Coverage}}                  \\
\midrule 
Case description & $\lambda$ range ($\rm \upmu m$)& Resolving power $res = \lambda/\delta\lambda$   \\
\midrule 
\textcolor{bluered_bandpass}{Blue \& Red}        & $[0.43 - 0.53]$ \& $[0.87 - 1.05]$  & 140 \& 140 \\
\textcolor{red_bandpass}{Red}                    & $[0.87 - 1.05]$                     & 140        \\
\textcolor{vis_bandpass}{Visible}                & $[0.45 - 1.00]$                     & 140        \\
\textcolor{NIR_bandpass}{NIR}                    & $[1.00-1.80]$                       & 70         \\
\textcolor{vNIR_bandpass}{vNIR} (Visible \& NIR) & $[0.45 - 1.00]$ \& $[1.00 - 1.80]$  & 140 \& 70  \\
\midrule 
\multicolumn{3}{c}{\textit{Spectral Quality$^*$}}                     \\
\midrule 
Moderate      &  \multicolumn{2}{r}{$\rm S/N_{\lambda_{0}}$ = 10} \\
Intermediate  &  \multicolumn{2}{r}{$\rm S/N_{\lambda_{0}}$ = 15} \\
Moderate-high &  \multicolumn{2}{r}{$\rm S/N_{\lambda_{0}}$ = 20} \\
\bottomrule 
\multicolumn{3}{l}{$^*$The S/N value is set at the lower bound of each spectral coverage}\\
\multicolumn{3}{l}{\phantom{$^*$}(e.g., $\rm \lambda_0 = 0.45~\rm \upmu m$ for the Visible and vNIR ranges; see \autoref{sec:appendix_SNR}).}\\
\multicolumn{3}{l}{\phantom{$^*$}See \autoref{fig:spectrum_SNR} for the corresponding S/N as a function of wavelength.}\\
\end{tabular}}
\end{table}

\section{Results} \label{sec:results}

We first introduce our parameter inference classification---which defines the levels of constraints or detection strengths based on inference results and posterior distribution shapes---and, then, present the retrieval results for observations across the different spectral bandpasses (and their combinations). These results are shown in \autoref{fig:post_dist_gas_abundances}, \autoref{fig:post_dist_surface_conditions}, \autoref{fig:post_dist_clouds}, \autoref{fig:post_dist_planetary_bulk_parameters}, and \autoref{fig:post_dist_orbital_parameters} for different parameter categories. Quantitative estimates are also provided in \autoref{tab:ret_results_SNR10}, \autoref{tab:ret_results_SNR15}, and \autoref{tab:ret_results_SNR20} for S/N = 10, 15, and 20, respectively.

\subsection{Parameter Inference Classification} \label{subsec:inference_classification}
Parameter inference accuracy is classified in 5 categories based on the shape of the posterior distribution and depending on the level of constraint obtained. Retrieved parameters can be any of the following (from highest to lowest accuracy):
\begin{itemize}
    \item Constrained (``\textcolor{green}{\textbf{C}}''): the fiducial value is retrieved unequivocally. The posterior distribution typically exhibits a marked peak without tails extending toward extreme values, with the fiducial value included within the 68\% confidence interval whose width is no larger than 1 log-unit (i.e., 1 order of magnitude) for log-scale prior ranges spanning at least 5 log-units, and no larger than 0.5 log-units otherwise. This is typically the case of the water vapor mixing ratio for observations in the visible or vNIR ranges from S/N = 15 (\autoref{fig:post_dist_gas_abundances}). For linear-scale prior ranges, the 68\% confidence interval width is contained within $\pm$20\% of the input value.
    \item Detected (``\textcolor{lime}{\textbf{D}}''): the parameter is loosely constrained. The posterior distribution is peaked without tails toward extreme values but with a 68\% confidence interval width larger than an order of magnitude \citep[``detection'' case of][]{Feng2018}. This is typically the case of the water vapor mixing ratio for observations in the visible or vNIR ranges at S/N = 10 (\autoref{fig:post_dist_gas_abundances}).
    \item Weakly detected (``\textcolor{yellow}{\textbf{WD}}''): a restricted range of values are favored, but extreme values are not excluded. The posterior distribution has a marked peak but a 68\% confidence interval width larger than an order of magnitude and also a substantial tail toward extreme values (``weak detection'' case of \citealp{Feng2018} and equivalent to the ``sensitivity limit'' case of \citealp{Konrad2022}). An illustrative example is the posterior distribution of the molecular oxygen mixing ratio for observations in the visible or vNIR ranges at S/N = 10 (\autoref{fig:post_dist_gas_abundances}).
    \item Limited by upper or lower bounds (``\textcolor{orange}{\textbf{L}}''): values above/below an upper/lower threshold are excluded, but the parameter can equally take any other value. The posterior distribution has no marked peak and is flat across a limited portion of the prior range (``upper limit'' type of \citealp{Konrad2022}). The posterior distributions of the methane mixing ratio exhibit a clear upper limit for observations in the vNIR at any S/N value (\autoref{fig:post_dist_gas_abundances}).
    \item Not constrained (``\textcolor{red}{\textbf{NC}}''): all parameter values are equally likely across the prior range, i.e., the parameter can take any value over the specified range. The posterior distribution is flat across the entire (or nearly) prior range (``nondetection'' case and ``unconstrained'' type of \citealp{Feng2018} and \citealp{Konrad2022}, respectively). The molecular nitrogen mixing ratio is typically not constrained in all retrieval scenarios presented here (\autoref{fig:post_dist_gas_abundances}).
\end{itemize}

\subsection{Gas Abundances}
\begin{figure}
    \centering
    \includegraphics[width=1.\textwidth, height=.9\textheight, keepaspectratio]{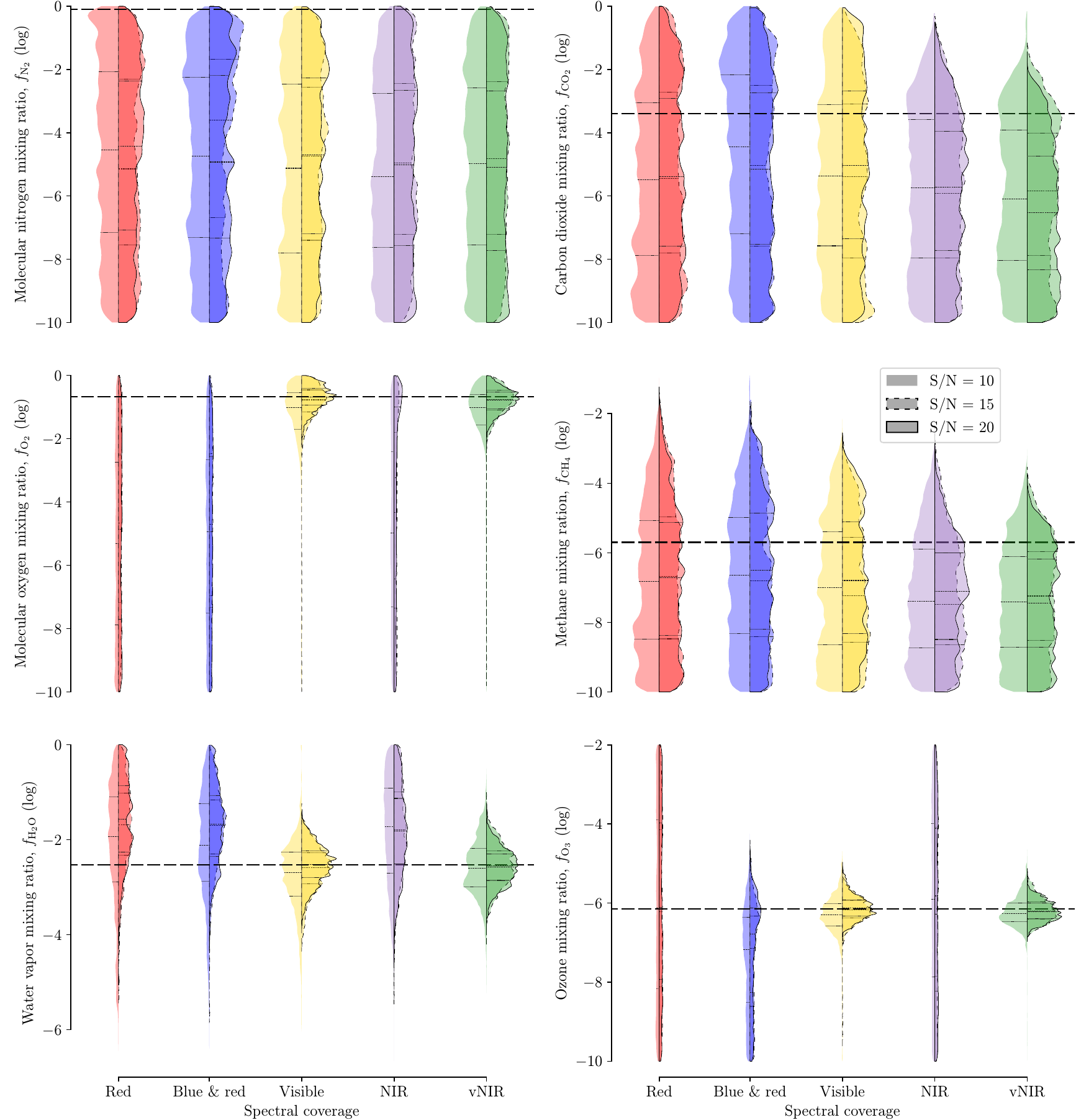}
    \caption{Posterior distributions of gas abundances obtained from observations in different color-coded spectral coverages at signal-to-noise ratios (S/N) of 10 (left side of the split-violins, without contour lines), 15 (right side, with dashed contours), and 20 (right side, with solid contours). From top to bottom: (left column) molecular nitrogen, molecular oxygen, water vapor, (right column) carbon dioxide, methane, and ozone. The horizontal dashed lines spanning each violin row indicate Earth-like input values, while the dashed lines within the violins show the 25\textsuperscript{th}, 50\textsuperscript{th}, and 75\textsuperscript{th} percentiles of the data (the first quartile, median, and third quartile, respectively). The density has been normalized across all kernel density plots so that each has the same area.
    }
    \label{fig:post_dist_gas_abundances}
\end{figure}

\autoref{fig:post_dist_gas_abundances} shows the retrieval results of the atmospheric composition.
Gas abundance retrieval accuracy generally correlates with the presence and strength of absorption and spectral features within the observed spectral bandpass (\autoref{fig:ref_spectrum}).
Prominent absorption features exist in each spectral range (except in the Blue; \autoref{fig:ref_spectrum}) and enable retrievals to provide at least lower limits on the atmospheric water vapor mixing ratio, regardless of the S/N (\autoref{fig:post_dist_gas_abundances}). For spectra of moderate quality ($\rm S/N=10$), a water vapor mixing ratio of at least $f_{\rm \ce{H2O}}\gtrsim 10^{-6}$ is necessary to explain the observed spectrum and its water vapor absorption features in the Red, Blue \& Red, and NIR bandpasses.
However, for these spectral coverages, further increases in S/N do not significantly improve inferences of water vapor concentration due to degeneracies with the abundance of other absorbers and cloud properties---especially cloud altitude (e.g., high-altitude clouds could make up for lower water vapor concentrations)---in the NIR (\autoref{fig:corner_NIR_SNR20}).
In contrast, firm detections of water vapor concentration can be achieved even at $\rm S/N = 10$ for observations covering broad spectral ranges in the visible (i.e., visible-only and vNIR), which benefit from constraints on other parameters---such as ozone and oxygen concentrations---that help break degeneracies (\autoref{fig:corner_vis_SNR10}). Furthermore, the water vapor mixing ratio can be constrained at an S/N of 15 for these spectral coverages.

Similar reasoning applies to the retrievals of ozone and molecular oxygen abundances, whose primary spectral features are found in the Visible bandpass, with minor absorption features for oxygen in the NIR.
As a result, the ozone mixing ratio can be constrained from observations spanning the visible (i.e., Visible and vNIR) at any S/N. An upper limit for $f_{\rm \ce{O3}}$ is obtained from observations in the combined Blue \& Red bandpass, but it cannot be constrained from the Red or NIR bandpasses, regardless of the S/N.
Oxygen has narrower spectral imprints than ozone, and due to the lack of significant spectral features in the red/blue bandpasses, it can only be weakly detected from observations in the Visible (owing mainly to the \ce{O2} $A$-band centered at 0.762 $\rm\upmu m$) and vNIR bandpasses. A minimum $\rm S/N = 20$ in the Visible and $\rm S/N = 15$ in the vNIR is required to fully constrain $f_{\rm \ce{O2}}$. It cannot be constrained from observations in the Red, NIR, or Blue \& Red bandpasses, regardless of the S/N.

Due to minor absorption features predominantly located in the NIR, retrievals can only place upper limits on the carbon dioxide abundance from observations in the NIR and vNIR bandpasses, with only marginal improvements in the latter. In the remaining spectral bandpasses analyzed, where \ce{CO2} is relatively transparent, $f_{\rm \ce{CO2}}$ cannot be constrained. Across all spectral coverages, increasing the S/N does not significantly improve the retrievals of $f_{\rm \ce{CO2}}$.
Regardless of the selected spectral bandpass, only upper limits can be placed on the methane abundance, with minor improvements from observations in the vNIR. Increasing the S/N does not significantly affect the retrievals of $f_{\rm \ce{CH4}}$. This is in agreement with previous studies showing that modern-Earth methane concentration is not detectable \citep[e.g.,][]{Kawashima2019, Latouf2025}.

Because of the absence of associated absorption features, the atmospheric abundance of molecular nitrogen can virtually take any value from $f_{\rm \ce{N2}} = 10^{-10}$ to $f_{\rm \ce{N2}} = 1$ (\autoref{fig:post_dist_gas_abundances}) and still produce a reflected-light spectrum that matches the observations. The lack of any apparent correlation with other parameters (e.g., \autoref{fig:corner_vis_SNR10}, \autoref{fig:corner_NIR_SNR20}) indicates that the atmosphere can be filled interchangeably with argon (the assumed background gas) or with molecular nitrogen. As a result, the abundance of molecular nitrogen cannot be constrained.

Overall, regarding the atmospheric composition, only $f_{\rm \ce{H2O}}$, $f_{\rm \ce{O3}}$, and $f_{\rm \ce{O2}}$ can be effectively constrained, with strong sensitivity to the spectral region observed and only a weak influence of the S/N. This agrees with the results of \citet{Susemiehl2023, Latouf2023a, Latouf2023b}, who used alternative restricted bandpasses. However, their more limited number of retrieved parameters may underestimate degeneracies that arise in retrievals with larger parameter spaces, potentially leading to artificially narrow posterior constraints. For the abundance of carbon dioxide and methane, only upper limits can be determined, while the abundance of \ce{N2} cannot be constrained in any of the observation scenarios.
We note that despite general agreement across recent retrieval studies \citep{Feng2018, Damiano2022, Susemiehl2023, Latouf2023a, Latouf2023b}, there are noticeable differences in the reported detection significances and the S/N thresholds required. For instance, while our study finds only weak detections of \ce{O2} at S/N = 10--15 in the visible range, \citet{Feng2018} report \ce{O2} detection in the same configuration (though defining S/N at a slightly different reference wavelength: 0.55 $\rm \upmu m$ compared to 0.45 $\rm \upmu m$ in the present study), and \citet{Latouf2023b} report strong \ce{O2} detections from S/N = 8 for 20\% width bandpasses centered between 0.68 and 0.84 $\rm \upmu m$. These discrepancies likely arise from a combination of factors, including differences in bandpass choices, S/N reference wavelengths and definitions, the number of free parameters, assumptions about background gas composition, MCMC convergence, and how detection significance is defined. A systematic intercomparison of retrieval assumptions and frameworks would be valuable to clarify these differences and establish more standardized benchmarks.

\subsection{Surface and Atmospheric Conditions}
\begin{figure}
    \centering
    \includegraphics[width=1.\textwidth, height=1.\textheight, keepaspectratio]{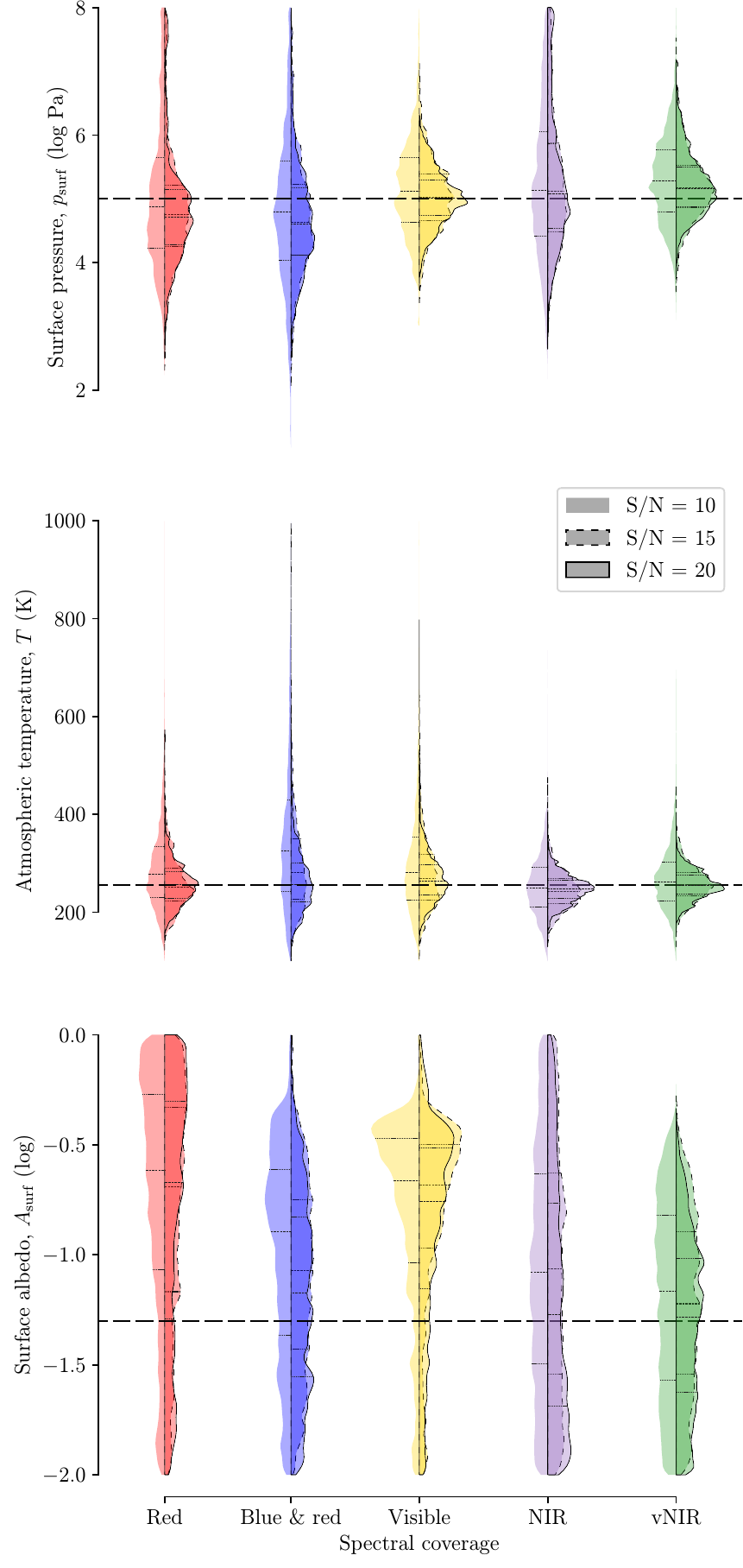}
    \caption{Posterior distributions of surface and atmospheric condition parameters obtained from observations in different color-coded spectral coverages at signal-to-noise ratios (S/N) of 10 (left side of the split-violins, without contour lines), 15 (right side, with dashed contours), and 20 (right side, with solid contours). From top to bottom: surface pressure, atmospheric temperature, and surface albedo. The horizontal dashed lines spanning each violin row indicate Earth-like input values, while the dashed lines within the violins show the 25\textsuperscript{th}, 50\textsuperscript{th}, and 75\textsuperscript{th} percentiles of the data (the first quartile, median, and third quartile, respectively). The density has been normalized across all kernel density plots so that each has the same area.}
    \label{fig:post_dist_surface_conditions}
\end{figure}

Both the surface pressure and atmospheric temperature are at least weakly detected for any combination of spectral coverage and S/N (\autoref{fig:post_dist_surface_conditions}). The surface pressure is best retrieved for observations including the visible range (i.e., Visible and vNIR): it is detected for S/N values of 10 and 15, and becomes constrained at S/N = 20.
For the characteristic atmospheric temperature, the best constraints come from observations including the NIR (i.e., vNIR and NIR), or part of it (i.e., Red). Both for observations restricted to the NIR or combined with the visible range, the temperature is detected at S/N = 10 and constrained from S/N = 15. It can be also detected and even constrained from observations in the Red bandpass at $\rm S/N = 15$ and $\rm S/N = 20$, respectively.
Conversely, the surface albedo is at most weakly detected, and requires observations covering as much visible range as possible to obtain upper limits. Increasing the S/N does not improve albedo characterization. The weak detections obtained for observations in the Visible bandpass come from the emergence of a peak in the posterior distribution. Yet, this peak systematically overestimates the surface albedo, favoring planets brighter than they actually are.

\subsection{Cloud Parameters}
\begin{figure}
    \centering
    \includegraphics[width=1.\textwidth, height=.95\textheight, keepaspectratio]{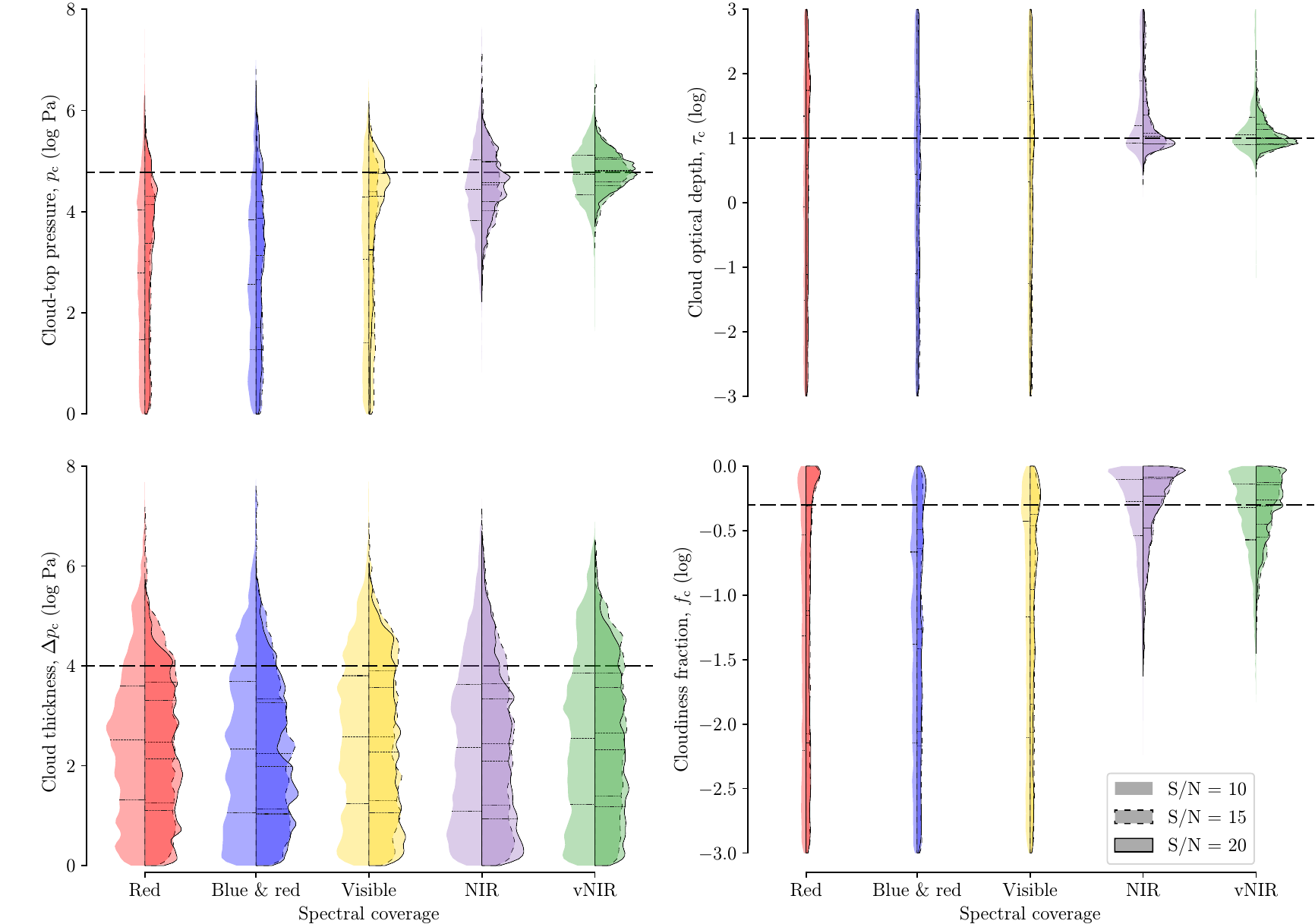}
    \caption{Posterior distributions of cloud properties obtained from observations in different color-coded spectral coverages at signal-to-noise ratios (S/N) of 10 (left side of the split-violins, without contour lines), 15 (right side, with dashed contours), and 20 (right side, with solid contours). From top to bottom: (left column) cloud-top pressure, thickness, (right column) optical depth, and cloudiness fraction. The horizontal dashed lines spanning each violin row indicate Earth-like input values, while the dashed lines within the violins show the 25\textsuperscript{th}, 50\textsuperscript{th}, and 75\textsuperscript{th} percentiles of the data (the first quartile, median, and third quartile, respectively). The density has been normalized across all kernel density plots so that each has the same area.
    }
    \label{fig:post_dist_clouds}
\end{figure}

Earth's realistic water ice and liquid clouds optical properties are parameterized with wavelength-dependent scattering properties and a Henyey-Greenstein phase function \citep{Robinson2023}. A cloud asymmetry parameter and single-scattering albedo are used to reproduce their spectral imprint in Earth's spectrum. Notably, Earth's water ice clouds absorb in the NIR, near 1.6 $\upmu$m \citep[e.g.,][]{Robinson2011}, which explains why removing all atmospheric species still leaves a pit relative to the continuum in that region (see \autoref{fig:ref_spectrum} and the differences between the fiducial reflected-light spectrum and the one assuming grey cloud properties). Because of this cloud optical properties-related spectral feature in the NIR, cloud optical depth, cloud-top pressure, and cloudiness fraction are always at least weakly detected for observations spanning the NIR, regardless of the S/N (\autoref{fig:post_dist_clouds}).
While these parameters are not accurately retrieved for observations conducted solely in the visible range, combining the Visible with the NIR gives the most accurate estimations, as seen by the shrinking tails of the posterior distributions and their narrowing toward the fiducial values, which are increasingly improved with increasing S/N. Notably, observations in the vNIR allow the cloudiness fraction to be constrained from S/N = 20 (weakly detected otherwise), the cloud-top pressure from S/N = 15 (already detected at S/N = 10), and the cloud optical depth from S/N = 10.
In contrast, retrievals of cloud thickness are not strongly sensitive to the observed spectral region or S/N---although increasing the S/N to 20 slightly refines the upper boundary---and only very thick clouds can be excluded.

\subsection{Planetary Bulk Parameters and Orbital Parameters}
\begin{figure}
    \centering
    \includegraphics[width=1.\textwidth, height=.4\textheight, keepaspectratio]{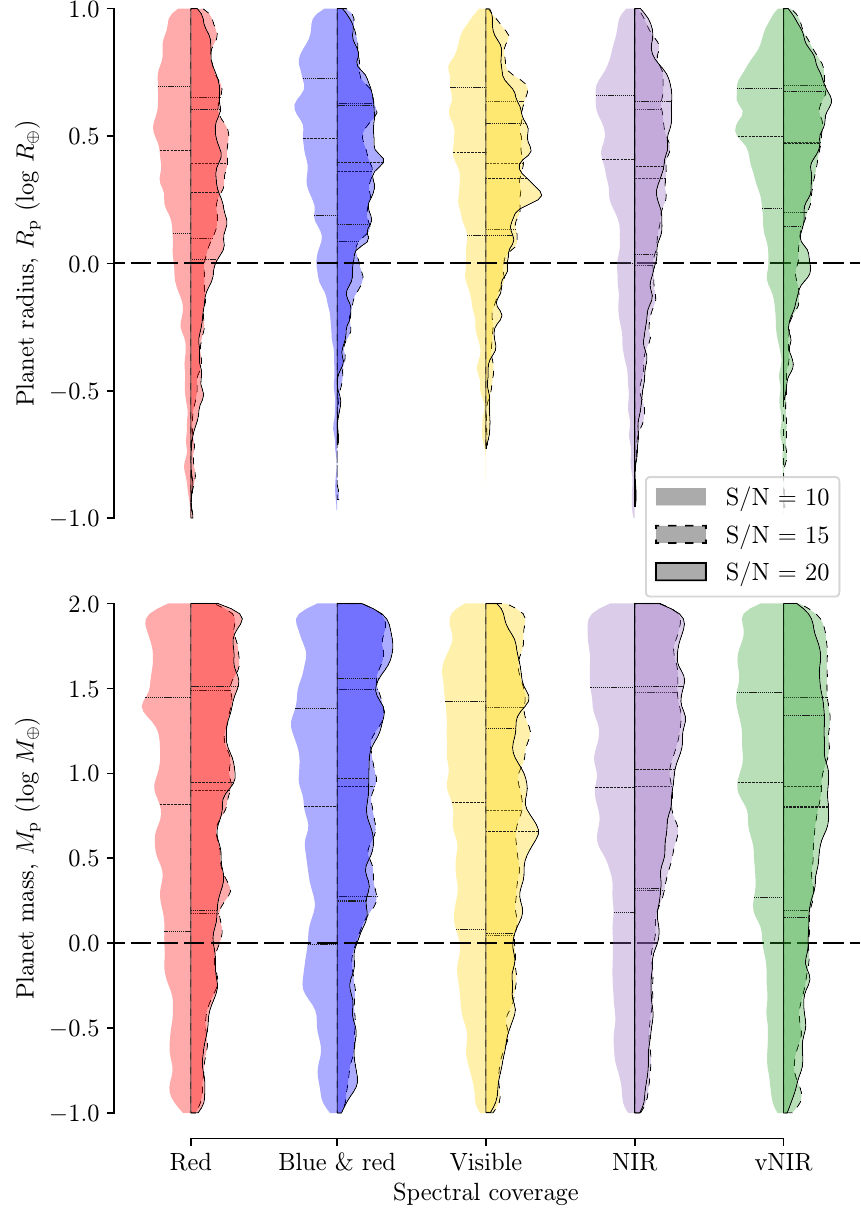}
    \caption{Posterior distributions of the planet radius (top) and mass (bottom) obtained from observations in different color-coded spectral coverages at signal-to-noise ratios (S/N) of 10 (left side of the split-violins, without contour lines), 15 (right side, with dashed contours), and 20 (right side, with solid contours). The horizontal dashed lines spanning each violin row indicate Earth-like input values, while the dashed lines within the violins show the 25\textsuperscript{th}, 50\textsuperscript{th}, and 75\textsuperscript{th} percentiles of the data (the first quartile, median, and third quartile, respectively). The density has been normalized across all kernel density plots so that each has the same area.}
    \label{fig:post_dist_planetary_bulk_parameters}
\end{figure}

Retrievals of the planetary bulk parameters are not sensitive to the spectral coverage or S/N (\autoref{fig:post_dist_planetary_bulk_parameters}).
Indeed, regardless of the observed spectral region or S/N, the planetary radius can only be weakly detected and is consistently overestimated. Observations in the Visible bandpass provide the best, though still unsatisfactory, lower limit. Note that \citet{Salvador2024} recently demonstrated that prior knowledge of the orbital parameters would allow tight constraints on the planetary radius.
In contrast, the planetary mass cannot be constrained in any of the proposed scenarios.

\begin{figure}
    \centering
    \includegraphics[width=1.\textwidth, height=.4\textheight, keepaspectratio]{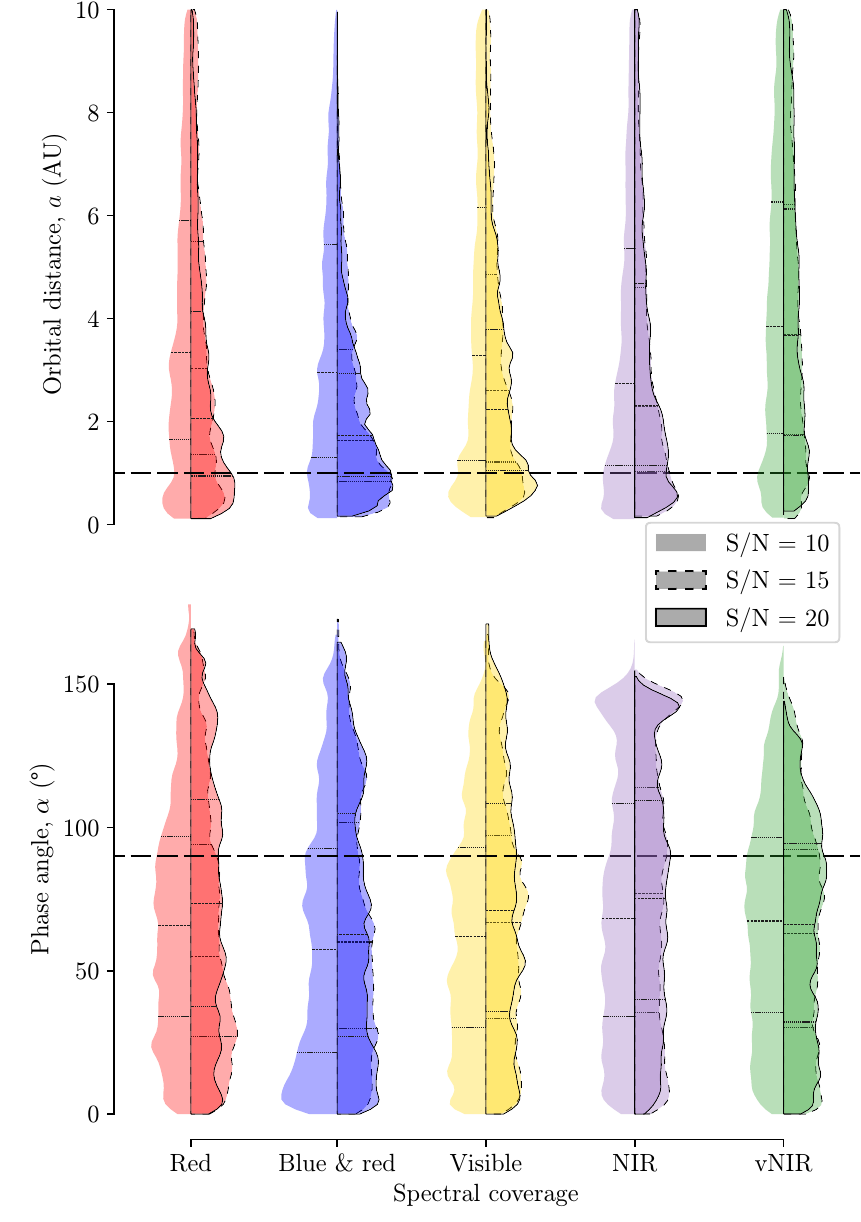}
    \caption{Posterior distributions of the orbital distance (top) and planetary phase angle (bottom) obtained from observations in different color-coded spectral coverages at signal-to-noise ratios (S/N) of 10 (left side of the split-violins, without contour lines), 15 (right side, with dashed contours), and 20 (right side, with solid contours). The horizontal dashed lines spanning each violin row indicate Earth-like input values, while the dashed lines within the violins show the 25\textsuperscript{th}, 50\textsuperscript{th}, and 75\textsuperscript{th} percentiles of the data (the first quartile, median, and third quartile, respectively). The density has been normalized across all kernel density plots so that each has the same area.}
    \label{fig:post_dist_orbital_parameters}
\end{figure}

The orbital parameters are also poorly constrained from reflected-light observations (\autoref{fig:post_dist_orbital_parameters}).
Due to their strong degeneracy with the planetary radius \citep[as demonstrated by][]{Nayak2017, Salvador2024}, and in the absence of any constraint on the latter, only a rough upper limit can be placed on the planetary phase angle. The orbital distance is at most weakly detected, with its posterior distribution peaking towards Earth-like values.

\section{Discussion} \label{sec:discuss}

Based on the retrieval results described above, \autoref{tab:constraint_levels} shows the inference level obtained for each retrieved parameter as a function of the spectral coverage and S/N, and \autoref{tab:min_requirements} summarizes the minimum observation requirements (in terms of spectral coverage and S/N) that HWO needs to meet to achieve the best possible level of constraint for each retrieved parameter.
Reciprocally, the relative assets of each spectral coverage and S/N combination are shown in \autoref{tab:detection_capabilities} and discussed in \autoref{subsec:detection_capabilities}. In \autoref{subsec:assessment_spectrum_first}, we assess the value of early spectral characterization prior to orbit determination, and in \autoref{subsec:observatory_architectures}, we discuss the implications of our findings for future observatory architectures.

\subsection{Observation Requirements for the Retrieved Parameters}
Summarizing our results, \autoref{tab:constraint_levels} and \autoref{tab:min_requirements} show that from the atmospheric composition, only \ce{H2O}, \ce{O2}, and \ce{O3} can be constrained and require observations spanning at least the Visible bandpass. In the absence of retrieval degeneracies when a large set of parameters are inferred or if more parameters are constrained, this could possibly be achieved with alternative restricted bandpasses than the ones presented here \citep{Susemiehl2023, Latouf2023a, Latouf2023b}.
Complementary observations in the mid-infrared would provide additional and missing constraints on the \ce{CO2} and \ce{CH4} abundances \citep[e.g.,][]{Konrad2022, Alei2024}, thus capturing a comprehensive picture of the atmospheric composition.

Except for the cloud thickness that is only limited and underestimated, cloud properties can be at least weakly detected from the NIR, and up to constrained when it is combined with the Visible. Regarding the surface conditions, the surface pressure is always at least weakly detected. It is detected for observations spanning the visible band (i.e., Visible and vNIR) at moderate (S/N = 10) and intermediate spectral quality (S/N = 15) and is constrained from S/N = 20.
The characteristic atmospheric temperature is at least weakly detected in the Visible, Blue \& Red, and Red bandpass cases, and becomes detected and constrained with increasing S/N in the latter. When the NIR is included, the temperature value is detected from moderate quality (S/N = 10) observations, and accurately constrained from S/N = 15.
Surface albedo inference can only be achieved with observations spanning a broad optical range, but is at most weakly detected and overestimated from observations in the Visible only. The most realistic estimates come from observations in the vNIR, but they only provide an upper limit.

The planetary radius is weakly detected at best and is typically overestimated, mistakenly favoring Neptune-sized planets. Prior information on the orbit \citep{Salvador2024}, mid-infrared observations \citep{Konrad2022}, or upcoming transit surveys \citep[such as PLATO;][]{Rauer2014, Rauer2016, Rauer2024plato} could help address this limitation, enabling more precise radius measurements -- although both direct imaging and transit observations would only be feasible for a restricted subset of targets \citep{Stark2020}.
Extreme Precision Radial Velocity could constrain the planetary mass \citep[e.g.,][]{Plavchan2015}, which is never constrained by the reflected-light observation scenarios proposed here. Similarly, the orbital parameters will require multiple reflected light observations over the course of the orbit, transit photometry, radial velocity measurements or astrometry to be more than weakly detected (in the Blue \& Red, Visible, and NIR cases, and at high S/N in the Red bandpass) for the orbital distance, and better than limited (in all cases except for low S/N observations in the Red) for the planetary phase angle.

\begin{table}[h!]
\caption{Observation Requirements for Each Retrieved Parameter}
\label{tab:constraint_levels}
\centering
\midsepremove
\begin{tabular}{ c | c | @{} c @{} | @{} c @{} | @{}c@{} | @{}c@{} | @{}c@{} } 
\toprule 
Parameter     & S/N &    ~~~~ \textbf{\textcolor{red_bandpass}{Red}} ~~~~~  &    ~~ \textbf{\textcolor{bluered_bandpass}{Blue \& Red}} ~~~    &    ~~~~ \textbf{\textcolor{vis_bandpass}{Visible}} ~~~~~    &    ~~~~ \textbf{\textcolor{NIR_bandpass}{NIR}} ~~~~~    &    ~~~~ \textbf{\textcolor{vNIR_bandpass}{vNIR}} ~~~~~    \\
\midrule[0.9pt] 
\multirow{3}{*}{$f_{\rm N_2}$}  & 10 & \cellcolor{red!30}NC & \cellcolor{red!30}NC & \cellcolor{red!30}NC           & \cellcolor{red!30}NC      & \cellcolor{red!30}NC    \\
                                \cmidrule{2-7} 
                                & 15 & \cellcolor{red!30}NC & \cellcolor{red!30}NC & \cellcolor{red!30}NC           & \cellcolor{red!30}NC      & \cellcolor{red!30}NC    \\
                                \cmidrule{2-7} 
                                & 20 & \cellcolor{red!30}NC & \cellcolor{red!30}NC & \cellcolor{red!30}NC           & \cellcolor{red!30}NC      & \cellcolor{red!30}NC    \\
\midrule[0.9pt] 
\multirow{3}{*}{$f_{\rm O_2}$}  & 10 & \cellcolor{red!30}NC & \cellcolor{red!30}NC & \cellcolor{yellow!30}WD        & \cellcolor{red!30}NC      & \cellcolor{yellow!30}WD \\
                                \cmidrule{2-7} 
                                & 15 & \cellcolor{red!30}NC & \cellcolor{red!30}NC & \cellcolor{yellow!30}WD        & \cellcolor{red!30}NC      & \cellcolor{green!30}C   \\
                                \cmidrule{2-7} 
                                & 20 & \cellcolor{red!30}NC & \cellcolor{red!30}NC & \cellcolor{green!30}C          & \cellcolor{red!30}NC      & \cellcolor{green!30}C   \\
\midrule[0.9pt] 
\multirow{3}{*}{$f_{\rm H_2O}$} & 10 & \cellcolor{orange!30}L & \cellcolor{orange!30}L & \cellcolor{lime!30}D       & \cellcolor{orange!30}L    & \cellcolor{lime!30}D    \\
                                \cmidrule{2-7} 
                                & 15 & \cellcolor{orange!30}L & \cellcolor{orange!30}L & \cellcolor{green!30}C      & \cellcolor{orange!30}L    & \cellcolor{green!30}C   \\
                                \cmidrule{2-7} 
                                & 20 & \cellcolor{orange!30}L & \cellcolor{orange!30}L & \cellcolor{green!30}C      & \cellcolor{orange!30}L    & \cellcolor{green!30}C   \\
\midrule[0.9pt] 
\multirow{3}{*}{$f_{\rm CO_2}$} & 10 & \cellcolor{red!30}NC & \cellcolor{red!30}NC & \cellcolor{red!30}NC           & \cellcolor{orange!30}L    & \cellcolor{orange!30}L  \\
                                \cmidrule{2-7} 
                                & 15 & \cellcolor{red!30}NC & \cellcolor{red!30}NC & \cellcolor{red!30}NC           & \cellcolor{orange!30}L    & \cellcolor{orange!30}L  \\
                                \cmidrule{2-7} 
                                & 20 & \cellcolor{red!30}NC & \cellcolor{red!30}NC & \cellcolor{red!30}NC           & \cellcolor{orange!30}L    & \cellcolor{orange!30}L  \\
\midrule[0.9pt] 
\multirow{3}{*}{$f_{\rm CH_4}$} & 10 & \cellcolor{orange!30}L & \cellcolor{orange!30}L & \cellcolor{orange!30}L     & \cellcolor{orange!30}L    & \cellcolor{orange!30}L  \\
                                \cmidrule{2-7} 
                                & 15 & \cellcolor{orange!30}L & \cellcolor{orange!30}L & \cellcolor{orange!30}L     & \cellcolor{orange!30}L    & \cellcolor{orange!30}L  \\
                                \cmidrule{2-7} 
                                & 20 & \cellcolor{orange!30}L & \cellcolor{orange!30}L & \cellcolor{orange!30}L     & \cellcolor{orange!30}L    & \cellcolor{orange!30}L  \\
\midrule[0.9pt] 
\multirow{3}{*}{$f_{\rm O_3}$}  & 10 & \cellcolor{red!30}NC & \cellcolor{orange!30}L & \cellcolor{green!30}C        & \cellcolor{red!30}NC      & \cellcolor{green!30}C   \\
                                \cmidrule{2-7} 
                                & 15 & \cellcolor{red!30}NC & \cellcolor{orange!30}L & \cellcolor{green!30}C        & \cellcolor{red!30}NC      & \cellcolor{green!30}C   \\
                                \cmidrule{2-7} 
                                & 20 & \cellcolor{red!30}NC & \cellcolor{orange!30}L & \cellcolor{green!30}C        & \cellcolor{red!30}NC      & \cellcolor{green!30}C   \\
\midrule[0.9pt] 
\multirow{3}{*}{$p_{\rm surf}$} & 10 & \cellcolor{yellow!30}WD & \cellcolor{yellow!30}WD & \cellcolor{lime!30}D     & \cellcolor{yellow!30}WD   & \cellcolor{lime!30}D    \\
                                \cmidrule{2-7} 
                                & 15 & \cellcolor{yellow!30}WD & \cellcolor{yellow!30}WD & \cellcolor{lime!30}D     & \cellcolor{yellow!30}WD   & \cellcolor{lime!30}D    \\
                                \cmidrule{2-7} 
                                & 20 & \cellcolor{yellow!30}WD & \cellcolor{yellow!30}WD & \cellcolor{green!30}C    & \cellcolor{yellow!30}WD   & \cellcolor{green!30}C   \\
\midrule[0.9pt] 
\multirow{3}{*}{$T$}            & 10 & \cellcolor{yellow!30}WD & \cellcolor{yellow!30}WD & \cellcolor{yellow!30}WD  & \cellcolor{lime!30}D      & \cellcolor{lime!30}D    \\
                                \cmidrule{2-7} 
                                & 15 & \cellcolor{lime!30}D & \cellcolor{yellow!30}WD & \cellcolor{yellow!30}WD     & \cellcolor{green!30}C     & \cellcolor{green!30}C   \\
                                \cmidrule{2-7} 
                                & 20 & \cellcolor{green!30}C & \cellcolor{yellow!30}WD & \cellcolor{yellow!30}WD    & \cellcolor{green!30}C     & \cellcolor{green!30}C   \\
\midrule[0.9pt] 
\multirow{3}{*}{$A_{\rm surf}$} & 10 & \cellcolor{red!30}NC & \cellcolor{orange!30}L & \cellcolor{yellow!30}WD      & \cellcolor{red!30}NC      & \cellcolor{orange!30}L  \\
                                \cmidrule{2-7} 
                                & 15 & \cellcolor{red!30}NC & \cellcolor{orange!30}L & \cellcolor{yellow!30}WD      & \cellcolor{red!30}NC      & \cellcolor{orange!30}L  \\
                                \cmidrule{2-7} 
                                & 20 & \cellcolor{red!30}NC & \cellcolor{orange!30}L & \cellcolor{yellow!30}WD      & \cellcolor{red!30}NC      & \cellcolor{orange!30}L  \\
\midrule[0.9pt] 
\multirow{3}{*}{$R_{\rm p}$}    & 10 & \cellcolor{yellow!30}WD & \cellcolor{yellow!30}WD & \cellcolor{yellow!30}WD  & \cellcolor{yellow!30}WD   & \cellcolor{yellow!30}WD \\
                                \cmidrule{2-7} 
                                & 15 & \cellcolor{yellow!30}WD & \cellcolor{yellow!30}WD & \cellcolor{yellow!30}WD  & \cellcolor{yellow!30}WD   & \cellcolor{yellow!30}WD \\
                                \cmidrule{2-7} 
                                & 20 & \cellcolor{yellow!30}WD & \cellcolor{yellow!30}WD & \cellcolor{yellow!30}WD  & \cellcolor{yellow!30}WD   & \cellcolor{yellow!30}WD \\
\midrule[0.9pt] 
\multirow{3}{*}{$M_{\rm p}$}    & 10 & \cellcolor{red!30}NC & \cellcolor{red!30}NC & \cellcolor{red!30}NC           & \cellcolor{red!30}NC      & \cellcolor{red!30}NC    \\
                                \cmidrule{2-7} 
                                & 15 & \cellcolor{red!30}NC & \cellcolor{red!30}NC & \cellcolor{red!30}NC           & \cellcolor{red!30}NC      & \cellcolor{red!30}NC    \\
                                \cmidrule{2-7} 
                                & 20 & \cellcolor{red!30}NC & \cellcolor{red!30}NC & \cellcolor{red!30}NC           & \cellcolor{red!30}NC      & \cellcolor{red!30}NC    \\
\midrule[0.9pt] 
\multirow{3}{*}{$p_{\rm c}$}    & 10 & \cellcolor{orange!30}L & \cellcolor{orange!30}L & \cellcolor{orange!30}L     & \cellcolor{yellow!30}WD   & \cellcolor{lime!30}D    \\
                                \cmidrule{2-7} 
                                & 15 & \cellcolor{orange!30}L & \cellcolor{orange!30}L & \cellcolor{orange!30}L     & \cellcolor{yellow!30}WD   & \cellcolor{green!30}C   \\
                                \cmidrule{2-7} 
                                & 20 & \cellcolor{orange!30}L & \cellcolor{orange!30}L & \cellcolor{yellow!30}WD    & \cellcolor{lime!30}D      & \cellcolor{green!30}C   \\
\midrule[0.9pt] 
\multirow{3}{*}{$\Delta p_{\rm c}$} & 10 & \cellcolor{orange!30}L & \cellcolor{orange!30}L & \cellcolor{orange!30}L & \cellcolor{orange!30}L    & \cellcolor{orange!30}L  \\
                                    \cmidrule{2-7} 
                                    & 15 & \cellcolor{orange!30}L & \cellcolor{orange!30}L & \cellcolor{orange!30}L & \cellcolor{orange!30}L    & \cellcolor{orange!30}L  \\
                                    \cmidrule{2-7} 
                                    & 20 & \cellcolor{orange!30}L & \cellcolor{orange!30}L & \cellcolor{orange!30}L & \cellcolor{orange!30}L    & \cellcolor{orange!30}L  \\
\midrule[0.9pt] 
\multirow{3}{*}{$\tau_{\rm c}$} & 10 & \cellcolor{red!30}NC & \cellcolor{red!30}NC & \cellcolor{red!30}NC           & \cellcolor{yellow!30}WD   & \cellcolor{green!30}C   \\
                                \cmidrule{2-7} 
                                & 15 & \cellcolor{red!30}NC & \cellcolor{red!30}NC & \cellcolor{red!30}NC           & \cellcolor{yellow!30}WD   & \cellcolor{green!30}C   \\
                                \cmidrule{2-7} 
                                & 20 & \cellcolor{red!30}NC & \cellcolor{red!30}NC & \cellcolor{red!30}NC           & \cellcolor{green!30}C     & \cellcolor{green!30}C   \\
\midrule[0.9pt] 
\multirow{3}{*}{$f_{\rm c}$}    & 10 & \cellcolor{red!30}NC & \cellcolor{red!30}NC & \cellcolor{red!30}NC           & \cellcolor{yellow!30}WD   & \cellcolor{yellow!30}WD \\
                                \cmidrule{2-7} 
                                & 15 & \cellcolor{red!30}NC & \cellcolor{red!30}NC & \cellcolor{red!30}NC           & \cellcolor{yellow!30}WD   & \cellcolor{yellow!30}WD \\
                                \cmidrule{2-7} 
                                & 20 & \cellcolor{red!30}NC & \cellcolor{red!30}NC & \cellcolor{red!30}NC           & \cellcolor{yellow!30}WD   & \cellcolor{green!30}C   \\
\midrule[0.9pt] 
\multirow{3}{*}{$a$}            & 10 & \cellcolor{red!30}NC & \cellcolor{yellow!30}WD & \cellcolor{yellow!30}WD     & \cellcolor{yellow!30}WD   & \cellcolor{red!30}NC    \\
                                \cmidrule{2-7} 
                                & 15 & \cellcolor{red!30}NC & \cellcolor{yellow!30}WD & \cellcolor{yellow!30}WD     & \cellcolor{yellow!30}WD   & \cellcolor{red!30}NC    \\
                                \cmidrule{2-7} 
                                & 20 & \cellcolor{yellow!30}WD & \cellcolor{yellow!30}WD & \cellcolor{yellow!30}WD  & \cellcolor{yellow!30}WD   & \cellcolor{red!30}NC    \\
\midrule[0.9pt] 
\multirow{3}{*}{$\alpha$}       & 10 & \cellcolor{red!30}NC & \cellcolor{orange!30}L & \cellcolor{orange!30}L       & \cellcolor{orange!30}L    & \cellcolor{orange!30}L  \\
                                \cmidrule{2-7} 
                                & 15 & \cellcolor{orange!30}L & \cellcolor{orange!30}L & \cellcolor{orange!30}L     & \cellcolor{orange!30}L    & \cellcolor{orange!30}L  \\
                                \cmidrule{2-7} 
                                & 20 & \cellcolor{orange!30}L & \cellcolor{orange!30}L & \cellcolor{orange!30}L     & \cellcolor{orange!30}L    & \cellcolor{orange!30}L  \\
\bottomrule 
\end{tabular}
\midsepdefault
\end{table}

\begin{table}[h]
\caption{Minimum Observation Requirements, i.e., Narrowest Spectral Coverage and Lowest S/N, to Achieve the Best Possible Inference of Atmospheric and Bulk Properties of Earth-Analogs}
\label{tab:min_requirements}
\centering
\resizebox{\textwidth}{!}{\begin{tabular}{l c c c}
\toprule 
Parameter     & Spectral Coverage            & S/N   & Best Inference \\
\midrule 
\multicolumn{4}{c}{\textit{Surface Conditions}}                                                       \\
Surface pressure, $p_{\rm surf}$  &  \textcolor{vis_bandpass}{Visible}  &  \textcolor{vis_bandpass}{20}     & \textcolor{green!70}{\textbf{Constraint}} \\
Atmospheric temperature, $T$      &  \textcolor{NIR_bandpass}{NIR}/\textcolor{red_bandpass}{Red}  &  \textcolor{NIR_bandpass}{15}/\textcolor{red_bandpass}{20} &  \textcolor{green!70}{\textbf{Constraint}}\\
Surface albedo, $A_{\rm surf}$    &  \textcolor{vis_bandpass}{Visible}       &  \textcolor{vis_bandpass}{10}     & \textcolor{yellow!70}{\textbf{Weak Detection}}$^*$ \\
\midrule 
\multicolumn{4}{c}{\textit{Gas Abundances}}                                   \\
Molecular nitrogen mixing ratio, $f_{\rm\ce{N2}}$  &  NA            &  NA    & \textcolor{red!70}{\textbf{No Constraint}} \\
Molecular oxygen mixing ratio, $f_{\rm\ce{O2}}$    &  \textcolor{vNIR_bandpass}{vNIR}/\textcolor{vis_bandpass}{Visible}  &  \textcolor{vNIR_bandpass}{15}/\textcolor{vis_bandpass}{20}  & \textcolor{green!70}{\textbf{Constraint}} \\
Water vapor mixing ratio, $f_{\rm\ce{H2O}}$        &  \textcolor{vis_bandpass}{Visible}       &  \textcolor{vis_bandpass}{15}    & \textcolor{green!70}{\textbf{Constraint}} \\
Carbon dioxide mixing ratio, $f_{\rm\ce{CO2}}$     &  \textcolor{NIR_bandpass}{NIR}            &  \textcolor{NIR_bandpass}{10}      & \textcolor{orange!70}{\textbf{Limit}} \\
Methane mixing ration, $f_{\rm\ce{CH4}}$           &  NA            &  NA     & \textcolor{orange!70}{\textbf{Limit}} \\
Ozone mixing ratio, $f_{\rm\ce{O3}}$               &  \textcolor{vis_bandpass}{Visible}       &  \textcolor{vis_bandpass}{10}    & \textcolor{green!70}{\textbf{Constraint}} \\
\midrule 
\multicolumn{4}{c}{\textit{Planetary Bulk Parameters}}                                                \\
Planet radius, $R_{\rm p}$  &  NA  &  NA  & \textcolor{yellow!70}{\textbf{Weak Detection}} \\
Planet mass, $M_{\rm p}$    &  NA  &  NA  & \textcolor{red!70}{\textbf{No Constraint}} \\
\midrule 
\multicolumn{4}{c}{\textit{Cloud Parameters}}                                                         \\
Cloud-top pressure, $p_{\rm c}$      &  \textcolor{vNIR_bandpass}{vNIR}  &  \textcolor{vNIR_bandpass}{15}  & \textcolor{green!70}{\textbf{Constraint}} \\
Cloud thickness, $\Delta p_{\rm c}$  &  NA    &  NA  & \textcolor{orange!70}{\textbf{Limit}} \\
Cloud optical depth, $\tau_{\rm c}$  &  \textcolor{vNIR_bandpass}{vNIR}/\textcolor{NIR_bandpass}{NIR}  &  \textcolor{vNIR_bandpass}{10}/\textcolor{NIR_bandpass}{20}  & \textcolor{green!70}{\textbf{Constraint}} \\
Cloudiness fraction, $f_{\rm c}$     &  \textcolor{vNIR_bandpass}{vNIR}  &  \textcolor{vNIR_bandpass}{20}  & \textcolor{green!70}{\textbf{Constraint}} \\
\midrule 
\multicolumn{4}{c}{\textit{Orbital Parameters}}                                                       \\
Planetary orbital distance, $a$ &  \textcolor{red_bandpass}{Red}/\textcolor{bluered_bandpass}{Blue \& Red}/\textcolor{vis_bandpass}{Visible}/\textcolor{NIR_bandpass}{NIR}  &  \textcolor{red_bandpass}{20}/\textcolor{bluered_bandpass}{10}/\textcolor{vis_bandpass}{10}/\textcolor{NIR_bandpass}{10}  & \textcolor{yellow!70}{\textbf{Weak Detection}} \\
Planetary phase angle, $\alpha$ &  \textcolor{red_bandpass}{Red}/\textcolor{bluered_bandpass}{Blue \& Red}/\textcolor{vis_bandpass}{Visible}/\textcolor{NIR_bandpass}{NIR}  &  \textcolor{red_bandpass}{15}/\textcolor{bluered_bandpass}{10}/\textcolor{vis_bandpass}{10}/\textcolor{NIR_bandpass}{10}  & \textcolor{orange!70}{\textbf{Limit}} \\
\bottomrule 
\multicolumn{4}{l}{\textbf{Notes.}}\\
\multicolumn{4}{l}{See \autoref{subsec:inference_classification} for the inference classification.}\\
\multicolumn{4}{l}{Not Applicable (NA) for parameters whose inference does not change with the spectral coverage or S/N.}\\
\multicolumn{4}{l}{$^*$Overly bright planets are favored from observations in the Visible. While only placing an upper limit,}\\
\multicolumn{4}{l}{\phantom{$^*$}vNIR observations provide better surface albedo inferences (see \autoref{fig:post_dist_surface_conditions} and \autoref{subsec:detection_capabilities}).}
\end{tabular}}
\end{table}

\subsection{Detection Capabilities of the Different Observation Scenarios}\label{subsec:detection_capabilities}
\autoref{tab:detection_capabilities} provides the detection capabilities of each spectral coverage and S/N combination for the proposed observation scenarios.

Observations through coronagraph-restricted spectral bandpasses (i.e., Red and Blue \& Red) provide little information about the planet atmospheric and bulk properties.
The atmospheric composition cannot be retrieved as atmospheric abundance constraints are at most limited (only for \ce{H2O} and \ce{CH4}). This is also true for NIR-only observations.
It should be noted that despite being centered on one of its absorption bands, observations in the Red bandpass at most limit the possible range of water vapor abundance. However, assuming alternative bandpass centers with restricted bandwidth could yield better results and allow the detection of \ce{H2O}, \ce{O2}, and \ce{O3} \citep{Susemiehl2023, Latouf2023a, Latouf2023b}.
Apart from the detection of the characteristic atmospheric temperature at S/N = 15 and its constraint at S/N = 20 with the Red bandpass, coronagraph-restricted observations at best allow only weak detections of the surface pressure, temperature, planet radius, and orbital distance (only at S/N = 20 in the Red bandpass), and this, regardless of S/N.
Adding the Blue bandpass to the Red one improves ozone abundance inference by placing an upper limit and significantly restricting the range of possible values. With increasing S/N, the posterior distribution peaks towards the fiducial value, indicating that it could possibly be retrieved at higher S/N.
When adding the Blue bandpass, the inference of the surface albedo is also improved with a 68\% confidence interval shifting towards the fiducial value and an upper limit excluding overly bright planets favored with the Red bandpass.
Despite these improvements, the parameters are at most weakly detected in the Blue \& Red scenario.

The Visible range is critical for detecting key atmospheric species such as \ce{H2O}, \ce{O2}, and \ce{O3}, and constraining their abundance when increasing the S/N.
None of the other spectral bandpasses alone achieve weak detection for any of the atmospheric component, even at S/N = 20.
This range also needs to be included to detect (S/N = 10 and 15) and constrain (at S/N = 20) the surface pressure, which can only be weakly detected otherwise.

NIR-only observations provide the strongest constraints on the characteristic atmospheric temperature even from low S/N, and are required to infer cloud properties. Indeed, the cloud fraction, optical depth, and cloud-top pressure can only be retrieved from observations including the NIR.

Combining the Visible and NIR (vNIR) thus provides the most comprehensive atmospheric characterization, by achieving constraints on key atmospheric species (i.e., \ce{H2O}, \ce{O2}, and \ce{O3}; obtained from the Visible), surface pressure and temperature, as well as inferring cloud properties (i.e., top pressure, optical depth, and fraction; from the NIR).
In addition, while adding the NIR to the Visible seems to degrade the surface albedo inference from our detection strength classification, the switch from an overestimated weak detection in the Visible to an upper limit in the vNIR actually favors values closer to the fiducial one, which is then included within the 68\% confidence interval.
However, it should be noted that the orbital distance inference becomes depreciated in the vNIR compared to Visible- or NIR-only observations (from weakly detected with a slight peak in the distribution to not constrained).

None of the spectral bandpass and S/N combinations considered here allow retrieval of the planetary bulk properties.
While the planetary mass is never constrained, the weak detection of the planet's radius mistakenly favors large planets over Earth-sized ones \citep[as emphasized in][]{Salvador2024}.
This highlights the importance of prior determination of the orbit (which cannot be achieved in any of the observation scenarios presented here) to break its degeneracy with the radius, thereby enabling accurate inference of the radius and correct identification of Earth-sized planets \citep{Salvador2024}.

\subsection{Assessment of the Value of Early Spectral Observations in Unconstrained Orbits}\label{subsec:assessment_spectrum_first}
Our results suggest that the identification and comprehensive characterization of a habitable environment---including the detection of key atmospheric species, potential biomarkers, and constraints on surface conditions ($P_{\rm surf}$ and $T$)---can be achieved even without prior knowledge of the planet's orbit, provided early spectral observations span the Visible range. Specifically, observations in this range enable inference of \ce{H2O}, \ce{O2}, and \ce{O3} abundances, as well as surface pressure.

Spanning the NIR is critical to infer cloud properties and atmospheric temperature. Combined with the Visible range, this spectral breadth maximizes the observation returns even at moderate S/N of 10.
Such broad spectral coverage allows inference of important parameters---including key atmospheric species (\ce{H2O}, \ce{O2}, and \ce{O3}), surface pressure, atmospheric temperature, cloud fraction, optical depth, and cloud-top pressure---demonstrating that much of the planet's thermo-chemical environment can be characterized prior to orbit determination.

However, even at moderate-to-high quality (S/N = 20), early spectral observations confined to coronagraph-restricted bandpasses (Red and Blue \& Red) can only place limits on the abundance of certain atmospheric species (\ce{H2O}, \ce{O3}, and \ce{CH4}) and weakly detect surface conditions. Such narrow bandpass observations are insufficient to provide firm constraints on the planet's habitable potential. Note that alternative restricted bandpasses have been shown to be promising \citep{Susemiehl2023, Latouf2023a, Latouf2023b}, but their detection capabilities should further be tested with a larger number of unconstrained parameters.

Given these considerations, a viable observation strategy is to perform early spectral observations, immediately following the initial photometric survey of a system, on any candidates with contrast ratios consistent with rocky planets. An additional consideration for prioritizing spectral observations in these systems could be a planet's projected separation from its host star, which is likely to be similar to its orbital semi-major axis \citep{Guimond2019}. It may be the case, though, that two photometric visits to the system spaced closely in time are required to ensure sources are co-moving. Information gained from these spectra can then prioritize targets for more detailed orbital and spectral follow-up, reducing the need for multiple photometric visits to lower-priority worlds. In the future, direct imaging yield calculation tools \citep[e.g.,][]{Stark2019, Morgan2022, Stark2024, Morgan2024} could assess if an ``early spectral characterization'' model achieves more opportunities for studying potentially Earth-like worlds.

Critically, in this ``spectrum first'' approach, the choice of spectral bandpass is more important than increasing S/N \citep[in agreement with previous studies considering different levels of prior constraints;][]{Damiano2022, Damiano2023, Susemiehl2023, Latouf2023a, Latouf2023b}. The ability to infer a planet's thermo-chemical environment and habitability potential depends more strongly on broad spectral coverage than on higher S/N observations confined to narrower bandpasses.

\begin{table}[h!]
\caption{Detection Capabilities as a Function of the Spectral Coverage and S/N}
\label{tab:detection_capabilities}
\centering
\midsepremove
\resizebox{\textwidth}{!}{
\begin{tabular}{ c | @{} c @{} | @{} c @{} | @{} c @{} | @{} c @{} }
\toprule 
Spectral Coverage & ~~~~Detection Strength~~~~ & S/N = 10  &  S/N = 15  &  S/N = 20  \\
\midrule[0.9pt] 
\multirow{5}{*}{\parbox{3cm}{\centering \textbf{\textcolor{red_bandpass}{Red}}\\ $\rm \lambda = [0.87-1.05]~\upmu m$}}               
                                   & \cellcolor{red!30}No Constraint
                                   & \cellcolor{red!30}~~~~$f_{\rm N_2}$, $f_{\rm O_2}$, $f_{\rm CO_2}$, $f_{\rm O_3}$, $A_{\rm surf}$, $M_{\rm p}$, $\tau_{\rm c}$, $f_{\rm c}$, $a$, $\alpha$~~~~
                                   & \cellcolor{red!30}~~~~$f_{\rm N_2}$, $f_{\rm O_2}$, $f_{\rm CO_2}$, $f_{\rm O_3}$, $A_{\rm surf}$, $M_{\rm p}$, $\tau_{\rm c}$, $f_{\rm c}$, $a$~~~~
                                   & \cellcolor{red!30}~~~~$f_{\rm N_2}$, $f_{\rm O_2}$, $f_{\rm CO_2}$, $f_{\rm O_3}$, $A_{\rm surf}$, $M_{\rm p}$, $\tau_{\rm c}$, $f_{\rm c}$~~~~
                                   \\
                                   \cmidrule{2-5} 
                                   & \cellcolor{orange!30}Limited
                                   & \cellcolor{orange!30}$f_{\rm H_2O}$, $f_{\rm CH_4}$, $p_{\rm c}$, $\Delta p_{\rm c}$
                                   & \cellcolor{orange!30}$f_{\rm H_2O}$, $f_{\rm CH_4}$, $p_{\rm c}$, $\Delta p_{\rm c}$, $\alpha$
                                   & \cellcolor{orange!30}$f_{\rm H_2O}$, $f_{\rm CH_4}$, $p_{\rm c}$, $\Delta p_{\rm c}$, $\alpha$
                                   \\
                                   \cmidrule{2-5} 
                                   & \cellcolor{yellow!30}Weak Detection
                                   & \cellcolor{yellow!30}$p_{\rm surf}$, $T$, $R_{\rm p}$
                                   & \cellcolor{yellow!30}$p_{\rm surf}$, $R_{\rm p}$
                                   & \cellcolor{yellow!30}$p_{\rm surf}$, $R_{\rm p}$, $a$
                                   \\
                                   \cmidrule{2-5} 
                                   & \cellcolor{lime!30}Detection
                                   & \cellcolor{lime!30}
                                   & \cellcolor{lime!30}$T$
                                   & \cellcolor{lime!30}
                                   \\
                                   \cmidrule{2-5} 
                                   & \cellcolor{green!30}Constraint
                                   & \cellcolor{green!30}
                                   & \cellcolor{green!30}
                                   & \cellcolor{green!30}$T$ \\
\midrule[0.9pt] 
\multirow{3}{*}{\parbox{3cm}{\centering \textbf{\textcolor{bluered_bandpass}{Blue \& Red}} \\ $\rm \lambda = [0.43-0.53]$~\&\\$\rm [0.87-1.05]~\upmu m$}}       
                                   & \cellcolor{red!30}No Constraint
                                   & \cellcolor{red!30}$f_{\rm N_2}$, $f_{\rm O_2}$, $f_{\rm CO_2}$, $M_{\rm p}$, $\tau_{\rm c}$, $f_{\rm c}$
                                   & \cellcolor{red!30}$f_{\rm N_2}$, $f_{\rm O_2}$, $f_{\rm CO_2}$, $M_{\rm p}$, $\tau_{\rm c}$, $f_{\rm c}$
                                   & \cellcolor{red!30}$f_{\rm N_2}$, $f_{\rm O_2}$, $f_{\rm CO_2}$, $M_{\rm p}$, $\tau_{\rm c}$, $f_{\rm c}$
                                   \\
                                   \cmidrule{2-5} 
                                   & \cellcolor{orange!30}Limited
                                   & \cellcolor{orange!30}$f_{\rm H_2O}$, $f_{\rm CH_4}$, $f_{\rm O_3}$, $A_{\rm surf}$, $p_{\rm c}$, $\Delta p_{\rm c}$, $\alpha$
                                   & \cellcolor{orange!30}$f_{\rm H_2O}$, $f_{\rm CH_4}$, $f_{\rm O_3}$, $A_{\rm surf}$, $p_{\rm c}$, $\Delta p_{\rm c}$, $\alpha$
                                   & \cellcolor{orange!30}$f_{\rm H_2O}$, $f_{\rm CH_4}$, $f_{\rm O_3}$, $A_{\rm surf}$, $p_{\rm c}$, $\Delta p_{\rm c}$, $\alpha$
                                   \\
                                   \cmidrule{2-5} 
                                   & \cellcolor{yellow!30}Weak Detection
                                   & \cellcolor{yellow!30}$p_{\rm surf}$, $T$, $R_{\rm p}$, $a$
                                   & \cellcolor{yellow!30}$p_{\rm surf}$, $T$, $R_{\rm p}$, $a$
                                   & \cellcolor{yellow!30}$p_{\rm surf}$, $T$, $R_{\rm p}$, $a$
                                   \\
\midrule[0.9pt] 
\multirow{5}{*}{\parbox{3cm}{\centering \textbf{\textcolor{vis_bandpass}{Visible}}\\ $\rm \lambda = [0.45-1.00]~\upmu m$}}           
                                   & \cellcolor{red!30}No Constraint
                                   & \cellcolor{red!30}$f_{\rm N_2}$, $f_{\rm CO_2}$, $M_{\rm p}$, $\tau_{\rm c}$, $f_{\rm c}$
                                   & \cellcolor{red!30}$f_{\rm N_2}$, $f_{\rm CO_2}$, $M_{\rm p}$, $\tau_{\rm c}$, $f_{\rm c}$
                                   & \cellcolor{red!30}$f_{\rm N_2}$, $f_{\rm CO_2}$, $M_{\rm p}$, $\tau_{\rm c}$, $f_{\rm c}$
                                   \\
                                   \cmidrule{2-5} 
                                   & \cellcolor{orange!30}Limited
                                   & \cellcolor{orange!30}$f_{\rm CH_4}$, $p_{\rm c}$, $\Delta p_{\rm c}$, $\alpha$
                                   & \cellcolor{orange!30}$f_{\rm CH_4}$, $p_{\rm c}$, $\Delta p_{\rm c}$, $\alpha$
                                   & \cellcolor{orange!30}$f_{\rm CH_4}$, $\Delta p_{\rm c}$, $\alpha$
                                   \\
                                   \cmidrule{2-5} 
                                   & \cellcolor{yellow!30}Weak Detection
                                   & \cellcolor{yellow!30}$f_{\rm O_2}$, $T$, $A_{\rm surf}$, $R_{\rm p}$, $a$
                                   & \cellcolor{yellow!30}$f_{\rm O_2}$, $T$, $A_{\rm surf}$, $R_{\rm p}$, $a$
                                   & \cellcolor{yellow!30}$T$, $A_{\rm surf}$, $R_{\rm p}$, $p_{\rm c}$, $a$
                                   \\
                                   \cmidrule{2-5} 
                                   & \cellcolor{lime!30}Detection
                                   & \cellcolor{lime!30}$f_{\rm H_2O}$, $p_{\rm surf}$
                                   & \cellcolor{lime!30}$p_{\rm surf}$
                                   & \cellcolor{lime!30}
                                   \\
                                   \cmidrule{2-5} 
                                   & \cellcolor{green!30}Constraint
                                   & \cellcolor{green!30}$f_{\rm O_3}$
                                   & \cellcolor{green!30}$f_{\rm H_2O}$, $f_{\rm O_3}$
                                   & \cellcolor{green!30}$f_{\rm O_2}$, $f_{\rm H_2O}$, $f_{\rm O_3}$, $p_{\rm surf}$
                                   \\
\midrule[0.9pt] 
\multirow{5}{*}{\parbox{3cm}{\centering \textbf{\textcolor{NIR_bandpass}{NIR}}\\ $\rm \lambda = [1.00-1.80]~\upmu m$}} 
                                   & \cellcolor{red!30}No Constraint
                                   & \cellcolor{red!30}$f_{\rm N_2}$, $f_{\rm O_2}$, $f_{\rm O_3}$, $A_{\rm surf}$, $M_{\rm p}$
                                   & \cellcolor{red!30}$f_{\rm N_2}$, $f_{\rm O_2}$, $f_{\rm O_3}$, $A_{\rm surf}$, $M_{\rm p}$
                                   & \cellcolor{red!30}$f_{\rm N_2}$, $f_{\rm O_2}$, $f_{\rm O_3}$, $A_{\rm surf}$, $M_{\rm p}$
                                   \\
                                   \cmidrule{2-5} 
                                   & \cellcolor{orange!30}Limited
                                   & \cellcolor{orange!30}$f_{\rm H_2O}$, $f_{\rm CO_2}$, $f_{\rm CH_4}$, $\Delta p_{\rm c}$, $\alpha$
                                   & \cellcolor{orange!30}$f_{\rm H_2O}$, $f_{\rm CO_2}$, $f_{\rm CH_4}$, $\Delta p_{\rm c}$, $\alpha$
                                   & \cellcolor{orange!30}$f_{\rm H_2O}$, $f_{\rm CO_2}$, $f_{\rm CH_4}$, $\Delta p_{\rm c}$, $\alpha$
                                   \\
                                   \cmidrule{2-5} 
                                   & \cellcolor{yellow!30}Weak Detection
                                   & \cellcolor{yellow!30}$p_{\rm surf}$, $R_{\rm p}$, $p_{\rm c}$, $\tau_{\rm c}$, $f_{\rm c}$, $a$
                                   & \cellcolor{yellow!30}$p_{\rm surf}$, $R_{\rm p}$, $p_{\rm c}$, $\tau_{\rm c}$, $f_{\rm c}$, $a$
                                   & \cellcolor{yellow!30}$p_{\rm surf}$, $R_{\rm p}$, $f_{\rm c}$, $a$
                                   \\
                                   \cmidrule{2-5} 
                                   & \cellcolor{lime!30}Detection
                                   & \cellcolor{lime!30}$T$
                                   & \cellcolor{lime!30}
                                   & \cellcolor{lime!30}$p_{\rm c}$
                                   \\
                                   \cmidrule{2-5} 
                                   & \cellcolor{green!30}Constraint
                                   & \cellcolor{green!30}
                                   & \cellcolor{green!30}$T$
                                   & \cellcolor{green!30}$T$, $\tau_{\rm c}$
                                   \\
\midrule[0.9pt] 
\multirow{5}{*}{\parbox{3cm}{\centering \textbf{\textcolor{vNIR_bandpass}{vNIR}}\\ $\rm \lambda = [0.45-1.80]~\upmu m$}}
                                   & \cellcolor{red!30}No Constraint
                                   & \cellcolor{red!30}$f_{\rm N_2}$, $M_{\rm p}$, $a$
                                   & \cellcolor{red!30}$f_{\rm N_2}$, $M_{\rm p}$, $a$
                                   & \cellcolor{red!30}$f_{\rm N_2}$, $M_{\rm p}$, $a$
                                   \\
                                   \cmidrule{2-5} 
                                   & \cellcolor{orange!30}Limited
                                   & \cellcolor{orange!30}$f_{\rm CO_2}$, $f_{\rm CH_4}$, $A_{\rm surf}$, $\Delta p_{\rm c}$, $\alpha$
                                   & \cellcolor{orange!30}$f_{\rm CO_2}$, $f_{\rm CH_4}$, $A_{\rm surf}$, $\Delta p_{\rm c}$, $\alpha$
                                   & \cellcolor{orange!30}$f_{\rm CO_2}$, $f_{\rm CH_4}$, $A_{\rm surf}$, $\Delta p_{\rm c}$, $\alpha$
                                   \\
                                   \cmidrule{2-5} 
                                   & \cellcolor{yellow!30}Weak Detection
                                   & \cellcolor{yellow!30}$f_{\rm O_2}$, $R_{\rm p}$, $f_{\rm c}$
                                   & \cellcolor{yellow!30}$R_{\rm p}$, $f_{\rm c}$
                                   & \cellcolor{yellow!30}$R_{\rm p}$
                                   \\
                                   \cmidrule{2-5} 
                                   & \cellcolor{lime!30}Detection
                                   & \cellcolor{lime!30}$f_{\rm H_2O}$, $p_{\rm surf}$, $T$, $p_{\rm c}$
                                   & \cellcolor{lime!30}$p_{\rm surf}$
                                   & \cellcolor{lime!30}
                                   \\
                                   \cmidrule{2-5} 
                                   & \cellcolor{green!30}Constraint
                                   & \cellcolor{green!30}$f_{\rm O_3}$, $\tau_{\rm c}$
                                   & \cellcolor{green!30}$f_{\rm O_2}$, $f_{\rm H_2O}$, $f_{\rm O_3}$, $T$, $p_{\rm c}$, $\tau_{\rm c}$
                                   & \cellcolor{green!30}$f_{\rm O_2}$, $f_{\rm H_2O}$, $f_{\rm O_3}$, $p_{\rm surf}$, $T$, $p_{\rm c}$, $\tau_{\rm c}$, $f_{\rm c}$
                                   \\
\bottomrule 
\end{tabular}
}
\midsepdefault
\end{table}

\subsection{Implications for Future Observatory Architectures}\label{subsec:observatory_architectures}

Our finding that wider spectral coverage plays a more decisive role than higher S/N in constraining key atmospheric and surface properties has important implications for the design of future direct imaging missions such as HWO.

First, extended wavelength coverage---particularly into the NIR---is essential for inferring cloud properties and atmospheric temperature. However, an exo-Earth target is more likely to fall within the coronagraph's inner working angle (IWA) at longer wavelengths, limiting access to features at these wavelengths for planets at small angular separations. Accessing the full NIR range therefore requires minimizing the IWA and/or increasing the telescope aperture. A larger aperture not only improves angular resolution but also enhances photon collection, providing secondary benefits for S/N and overall characterization efficiency.

Second, by emphasizing the importance of broad spectral coverage, our results highlight the value of coronagraph designs that support wide instantaneous bandpasses or enable simultaneous observations across different spectral regions. This capability could be achieved through single multi-channel instruments or multiple instruments operating in parallel.

Taken together, these results support the development of mission architectures that:
\begin{enumerate}
    \item minimize the coronagraph's IWA to enable full access to the visible and NIR ranges at relevant angular separations;
    \item maximize the instrument bandpass width or enable multi-band parallel observations;
    \item prioritize moderate-S/N observations across broad wavelength ranges over high-S/N spectroscopy in narrow bands.
\end{enumerate}
Such designs would maximize the scientific return of early spectral observations of Earth analogs in unconstrained orbits, enabling robust atmospheric characterization even before orbit determination.

\section{Conclusion} \label{sec:conclusion}

The most efficient observing strategies for searching for habitable Earth-analogs observed in reflected light with future space-based telescopes have yet to be fully defined. In this work, we explored how spectral coverage---particularly in coronagraph-restricted bandpasses---and the quality of direct imaging observations affect the ability to identify atmospheric and bulk properties of rocky exoplanets, assuming spectral observations are taken ``early'' (i.e., before any prior orbit measurements).

Using our atmospheric retrieval tool \texttt{rfast}, we evaluated the inference of 17 key atmospheric and planetary parameters for HWO-like observations covering various spectral bandpasses and spectral qualities ranging from moderate (S/N = 10) to moderate-high (S/N = 20).
We deduced the observation requirements, in terms of spectral bandpass and quality (S/N), needed to infer this broad range of atmospheric and bulk exoplanet properties, and reciprocally assessed the detection capabilities of the tested observation scenarios.
Our key findings are:
\begin{itemize}
    \item[--] Atmospheric composition cannot be reliably retrieved from a single Red or combined Blue \& Red coronagraph-restricted bandpasses;
    \item[--] Constraints on \ce{H2O}, \ce{O2}, and \ce{O3} require observations spanning at least the Visible range, which also provides the best constraints on surface pressure;
    \item[--] Cloud fraction, optical depth, and cloud-top pressure can be detected in the NIR;
    \item[--] Observations combining Visible and NIR provide the best constraints for key atmospheric species abundances (\ce{H2O}, \ce{O2}, and \ce{O3}), cloud properties (cloud fraction, optical depth, and cloud-top pressure), surface pressure, and atmospheric temperature;
    \item[--] Without prior knowledge, planet radius is systematically overestimated;
    \item[--] Early spectral observations in unconstrained orbits can still robustly constrain the planet's thermo-chemical environment if spanning the Visible and NIR;
    \item[--] A ``spectrum first'' approach could efficiently guide target prioritization and reduce repeated photometric observations of low-priority worlds;
    \item[--] The choice of spectral bandpass is more critical than increasing S/N for retrieving atmospheric and surface properties and maximizing scientific return.
\end{itemize}

Overall, our results indicate that future direct imaging missions targeting Earth-analog characterization (such as the Habitable Worlds Observatory) could prioritize observations spanning broad spectral coverage, including the visible and possibly extending to the NIR, even at moderate S/N (10), rather than focusing on higher quality observations confined to narrow coronagraph bandpasses.
For example, a single coronagraph-restricted channel centered on a water absorption band can only place lower limits on water vapor abundance and is insufficient for detection, even at S/N = 20 combined with a Blue bandpass. In contrast, vNIR observations enable comprehensive constraints on the thermo-chemical state of the atmosphere through the inference of atmospheric composition, surface conditions, and cloud properties from moderate spectral quality data (S/N = 10).
This approach not only maximizes the characterization potential from early observations in unconstrained orbits but also informs efficient mission planning and target selection, ultimately enhancing the search for habitable Earth-analogs.

\section*{Acknowledgments}
A.S. and T.D.R. gratefully acknowledge support from NASA's Habitable Worlds Program (No.~80NSSC20K0226) and NASA's Nexus for Exoplanet System Science Virtual Planetary Laboratory (No.~80NSSC18K0829). T.D.R. and A.S. also acknowledge support from the Cottrell Scholar Program administered by the Research Corporation for Science Advancement.
A.S. warmly thanks the Bibliothèque Universitaire du Campus Périgord (Université de Bordeaux) and its team members, particularly Marie-Christine Emery-Bonnet, Catherine Dupin, and Nathalie Vidal, for providing office space and offering their welcoming support. A.S. is also grateful to Moina Issoufi for her invaluable assistance with figure design.
Both authors thank an anonymous reviewer for their insightful and constructive comments.

\vspace{5mm}
\software{
\texttt{Astropy} \citep{astropy:2013, astropy:2018, astropy:2022},
\texttt{corner} \citep{Foreman-Mackey2016},
\texttt{emcee} \citep{Foreman-Mackey2013},
\texttt{matplotlib} \citep{matplotlib},
\texttt{NumPy} \citep{numpy},
\texttt{pandas} \citep{pandas},
\texttt{rfast} \citep{Robinson2023},
\texttt{seaborn} \citep{seaborn}
}

\appendix

\section{Signal-to-noise ratio as a function of wavelength} \label{sec:appendix_SNR}
To inform mission design and planning tools---such as coronagraph exposure time and yield calculators---and facilitate integration of our results, we provide the signal-to-noise ratio (S/N) as a function of wavelength for each spectral coverage region considered (\autoref{tab:retrieval_grid}).

Three values of $\rm S/N_{\lambda_0}$ (10, 15, and 20) are considered to represent different spectral quality scenarios (\autoref{tab:retrieval_grid}).
We define $\rm S/N_{\lambda_0}$ as the S/N set at the lower bound of each spectral coverage, denoted $\lambda_0$. The corresponding noise level, computed at $\lambda_0$ for a given $\rm S/N_{\lambda_0}$, is kept constant and propagated across the full bandpass (i.e., the noise is wavelength-independent).
As a result, the S/N varies with wavelength and decreases within absorption features, where the reflected flux is lower \citep[e.g.,][]{Feng2018, Robinson2023, Salvador2024}.

\begin{figure}
    \centering
    \includegraphics[width=1.\textwidth, height=.5\textheight, keepaspectratio]{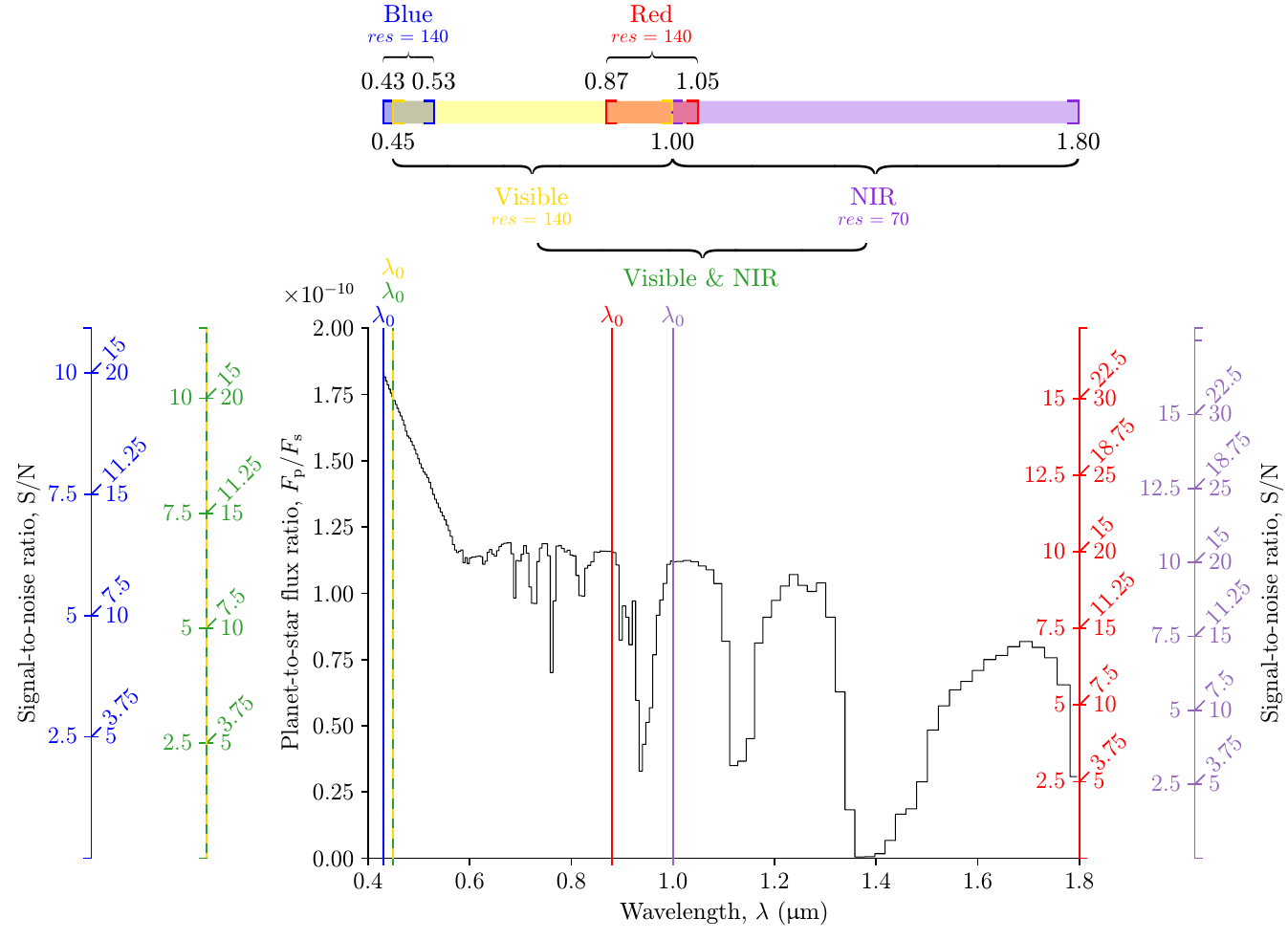}
    \caption{Reflected light spectrum of the fiducial case (black curve; see \autoref{fig:ref_spectrum} and \autoref{tab:retrieved_pars}) and corresponding signal-to-noise ratio (S/N; colored $y$-axes) as a function of wavelength for each color-coded spectral coverage (\autoref{tab:retrieval_grid}). The left, oblique, and right ticks of each colored $y$-axis correspond to $\rm S/N_{\lambda_0}=10$, 15, and 20, respectively, where $\lambda_0$ (indicated by vertical lines) is the lower bound of each spectral coverage at which the specified $\rm S/N_{\lambda_0}$ is defined. See text to extract the S/N at any wavelength of interest.
    }
    \label{fig:spectrum_SNR}
\end{figure}

\autoref{fig:spectrum_SNR} illustrates how the specified values $\rm S/N_{\lambda_0}=10$, 15, and 20 translate to S/N at any other wavelength for each spectral coverage.
For each color-coded spectral coverage and associated $y$-axis, the left, oblique, and right ticks correspond to $\rm S/N_{\lambda_0} = 10$, 15, and 20, respectively, allowing direct estimation of the S/N at any wavelength within the bandpass.

As an example, in the continuum near the 0.94~$\rm \upmu m$ water absorption feature, e.g., at $\lambda = 0.87~\rm \upmu m$, $\rm S/N_{0.87\,\upmu m}$ is approximately 6.5 and 13 for the vNIR spectral coverage (green) when $\rm S/N_{\lambda_0\,=\,0.45\,\rm \upmu m} = 10$ and 20, respectively. The same values apply to the Visible spectral coverage (yellow), which shares the same $\lambda_0$.

This enables estimation of the local S/N at specific spectral regions of interest, which is particularly relevant for determining the exposure time required by coronagraphs targeting key molecular features within relatively narrow bandpasses.

\section{Retrieval Results} \label{sec:appendix_retrieval_results}

\begin{figure}
    \centering
    \includegraphics[width=1.\textwidth, height=.95\textheight, keepaspectratio]{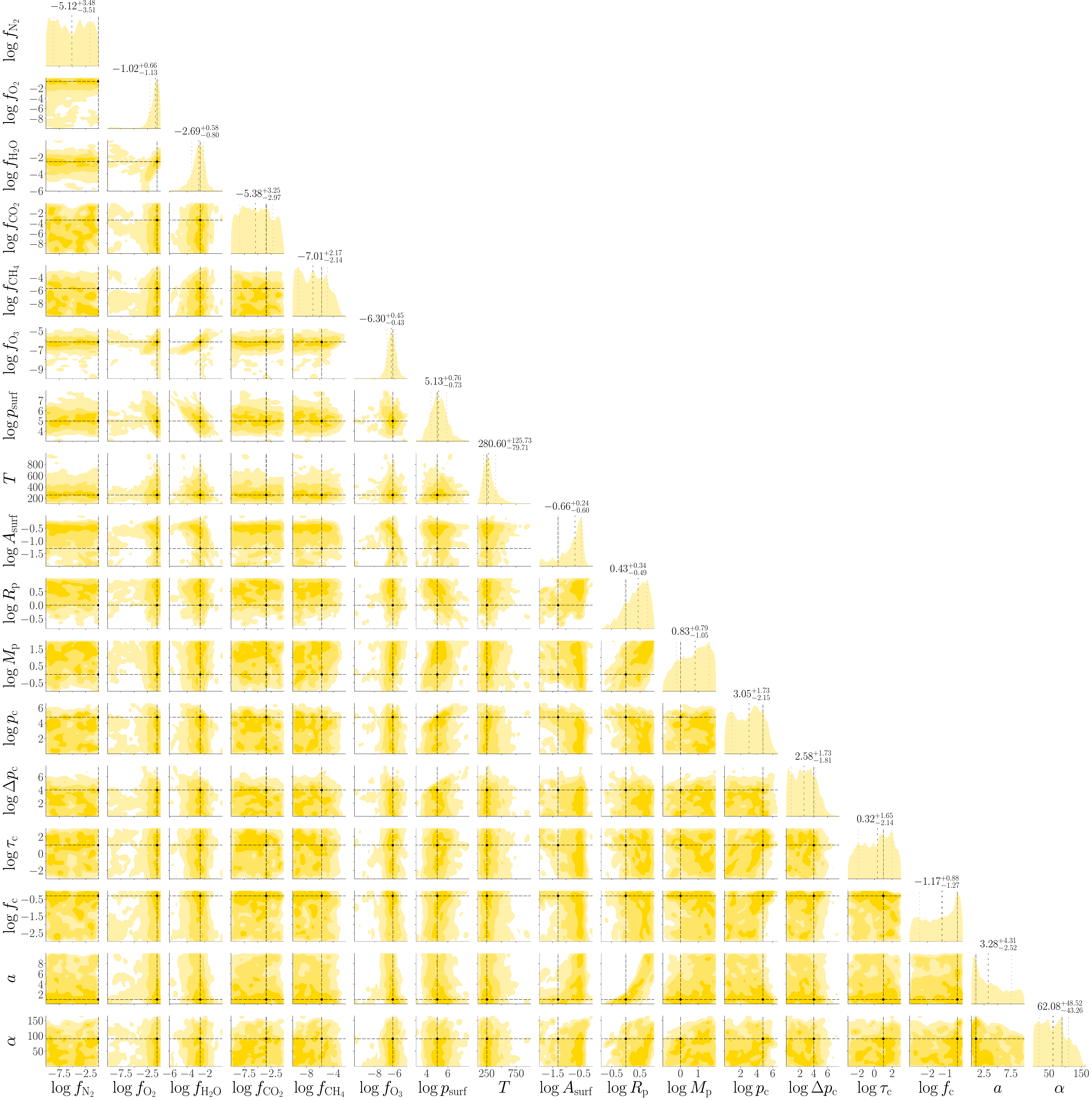}
    \caption{Corner plot showing the marginal univariate (along the diagonal) and joint bivariate (off-diagonal) posterior distributions of all 17 retrieved parameters from observations conducted in the Visible bandpass at $\rm S/N = 10$. The retrieved values (loosely dashed) and their $\pm1\sigma$ (68\%) credible intervals (loosely dotted) are indicated by vertical lines on the 1D marginal distributions (i.e., the 16\textsuperscript{th}, 50\textsuperscript{th}, and 84\textsuperscript{th} percentiles). The contours of the 2D posterior distributions correspond to the 1$\sigma$, 2$\sigma$, and 3$\sigma$ levels, which encompass 68\%, 95\%, and 99.7\% of the observed values, respectively. Note that these levels correspond to 39.3\%, 86.5\%, and 98.9\% of the volume in 2D distributions, and to 68\%, 95\%, and 99.7\% in 1D distributions. The vertical and horizontal dashed lines indicate the Earth-based input values.
    }
    \label{fig:corner_vis_SNR10}
\end{figure}

\begin{figure}
    \centering
    \includegraphics[width=1.\textwidth, height=.95\textheight, keepaspectratio]{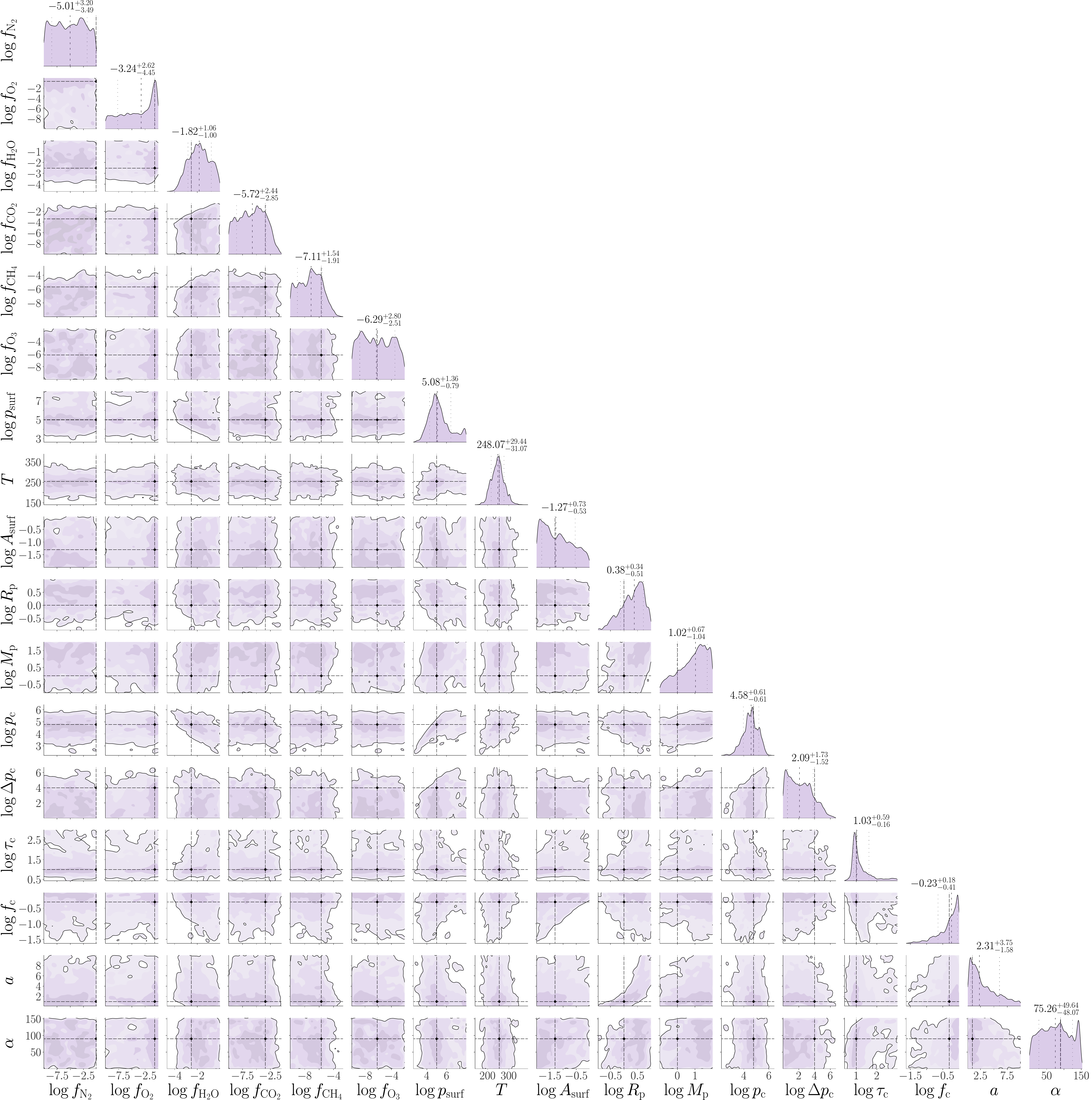}
    \caption{Corner plot showing the marginal univariate (along the diagonal) and joint bivariate (off-diagonal) posterior distributions of all 17 retrieved parameters from observations conducted in the NIR bandpass at $\rm S/N = 20$. The retrieved values (loosely dashed) and their $\pm1\sigma$ (68\%) credible intervals (loosely dotted) are indicated by vertical lines on the 1D marginal distributions (i.e., the 16\textsuperscript{th}, 50\textsuperscript{th}, and 84\textsuperscript{th} percentiles). The contours of the 2D posterior distributions correspond to the 1$\sigma$, 2$\sigma$, and 3$\sigma$ levels, which encompass 68\%, 95\%, and 99.7\% of the observed values, respectively. Note that these levels correspond to 39.3\%, 86.5\%, and 98.9\% of the volume in 2D distributions, and to 68\%, 95\%, and 99.7\% in 1D distributions. The vertical and horizontal dashed lines indicate the Earth-based input values.
    }
    \label{fig:corner_NIR_SNR20}
\end{figure}

Retrieval results obtained from observations conducted across the different spectral bandpasses are provided in:
\begin{itemize}
    \item[--] \autoref{tab:ret_results_SNR10} for $\rm S/N = 10$;
    \item[--] \autoref{tab:ret_results_SNR15} for $\rm S/N = 15$;
    \item[--] \autoref{tab:ret_results_SNR20} for $\rm S/N = 20$.
\end{itemize}
The Earth-based input values and the level of constraint for each parameter are also indicated (see \autoref{subsec:inference_classification} for the inference classification).
Corner plots showing the marginal univariate (along the diagonal) and joint bivariate (off-diagonal) posterior distributions of all retrieved parameters are provided in:
\begin{itemize}
    \item[--] \autoref{fig:corner_vis_SNR10} for observations in the Visible bandpass at $\rm S/N = 10$;
    \item[--] \autoref{fig:corner_NIR_SNR20} for observations in the NIR bandpass at $\rm S/N = 20$.
\end{itemize}


\begin{table}
\centering
\resizebox{1.\textwidth}{!}{%
\begin{threeparttable}
\caption{Retrieval Results Comparison and Associated Level of Constraint for Observations Conducted in Different Spectral Coverages at S/N = 10}\label{tab:ret_results_SNR10}
\begin{tabular}{lS|S@{\hspace{-0.17cm}}l@{}c|S@{\hspace{-0.17cm}}l@{}c|S@{\hspace{-0.17cm}}l@{}c|S@{\hspace{-0.17cm}}l@{}c|S@{\hspace{-0.17cm}}l@{}c|}
\toprule
Parameter & Input & \multicolumn{3}{c}{Red} & \multicolumn{3}{c}{Blue \& Red} & \multicolumn{3}{c}{Visible} & \multicolumn{3}{c}{NIR} & \multicolumn{3}{c}{vNIR} \\
 \midrule
$\log\,$$f_{\rm N_2}$         &   -0.11   &      -4.54  & $_{-3.62}^{+3.34}$ &  NC     &        -4.75  & $_{-3.50}^{+3.20}$ &  NC     &        -5.12  & $_{-3.51}^{+3.48}$ &  NC      &        -5.38  & $_{-3.10}^{+3.46}$ &  NC     &        -4.97  & $_{-3.50}^{+3.18}$ &  NC     \\
$\log\,$$f_{\rm O_2}$         &   -0.68   &       -5.3  & $_{-3.41}^{+3.38}$ &  NC     &        -4.94  & $_{-3.42}^{+3.20}$ &  NC     &        -1.02  & $_{-1.13}^{+0.66}$ &  WD      &        -4.99  & $_{-3.22}^{+3.47}$ &  NC     &        -1.02  & $_{-0.89}^{+0.59}$ &  WD     \\
$\log\,$$f_{\rm H_2O}$        &   -2.52   &      -1.94  & $_{-1.59}^{+1.13}$ &  L      &        -2.12  & $_{-1.17}^{+1.17}$ &  L      &        -2.69  & $_{-0.80}^{+0.58}$ &  D       &        -1.73  & $_{-1.39}^{+1.11}$ &  L      &         -2.6  & $_{-0.55}^{+0.65}$ &  D      \\
$\log\,$$f_{\rm CO_2}$        &    -3.4   &      -5.49  & $_{-3.26}^{+3.28}$ &  NC     &        -4.45  & $_{-3.77}^{+2.89}$ &  NC     &        -5.38  & $_{-2.97}^{+3.25}$ &  NC      &        -5.73  & $_{-2.93}^{+2.84}$ &  L      &         -6.1  & $_{-2.57}^{+2.90}$ &  L      \\
$\log\,$$f_{\rm CH_4}$        &    -5.7   &      -6.83  & $_{-2.18}^{+2.42}$ &  L      &        -6.65  & $_{-2.27}^{+2.33}$ &  L      &        -7.01  & $_{-2.14}^{+2.17}$ &  L       &        -7.39  & $_{-1.80}^{+2.07}$ &  L      &        -7.42  & $_{-1.76}^{+1.82}$ &  L      \\
$\log\,$$f_{\rm O_3}$         &   -6.15   &      -6.13  & $_{-2.71}^{+2.92}$ &  NC     &        -7.18  & $_{-1.83}^{+1.11}$ &  L      &         -6.3  & $_{-0.43}^{+0.45}$ &  C       &         -5.9  & $_{-2.77}^{+2.56}$ &  NC     &        -6.26  & $_{-0.31}^{+0.42}$ &  C      \\
$\log\,$$p_{\rm surf}$        &    5.0    &       4.88  & $_{-0.94}^{+1.44}$ &  WD     &          4.8  & $_{-1.11}^{+1.25}$ &  WD     &         5.13  & $_{-0.73}^{+0.76}$ &  D       &         5.13  & $_{-1.06}^{+1.43}$ &  WD     &         5.28  & $_{-0.70}^{+0.70}$ &  D      \\
$T$                           &   255.0   &     277.09  & $_{-66.23}^{+95.47}$ &  WD   &       324.76  & $_{-116.35}^{+186.54}$ &  WD &        280.6  & $_{-79.71}^{+125.73}$ &  WD   &       249.71  & $_{-58.48}^{+63.87}$ &  D    &       262.43  & $_{-55.46}^{+64.32}$ &  D    \\
$\log\,$$A_{\rm surf}$        &    -1.3   &      -0.62  & $_{-0.73}^{+0.44}$ &  NC     &        -0.89  & $_{-0.66}^{+0.40}$ &  L      &        -0.66  & $_{-0.60}^{+0.24}$ &  WD      &        -1.08  & $_{-0.59}^{+0.62}$ &  NC     &        -1.17  & $_{-0.55}^{+0.47}$ &  L      \\
$\log\,$$R_{\rm p}$           &    0.0    &       0.44  & $_{-0.51}^{+0.35}$ &  WD     &         0.49  & $_{-0.47}^{+0.32}$ &  WD     &         0.43  & $_{-0.49}^{+0.34}$ &  WD      &         0.41  & $_{-0.59}^{+0.33}$ &  WD     &          0.5  & $_{-0.45}^{+0.26}$ &  WD     \\
$\log\,$$M_{\rm p}$           &    0.0    &       0.82  & $_{-1.06}^{+0.84}$ &  NC     &          0.8  & $_{-1.07}^{+0.80}$ &  NC     &         0.83  & $_{-1.05}^{+0.79}$ &  NC      &         0.92  & $_{-1.12}^{+0.77}$ &  NC     &         0.95  & $_{-1.00}^{+0.71}$ &  NC     \\
$\log\,$$p_{\rm c}$           &    4.78   &       2.79  & $_{-1.83}^{+1.65}$ &  L      &         2.56  & $_{-1.76}^{+1.74}$ &  L      &         3.05  & $_{-2.15}^{+1.73}$ &  L       &         4.44  & $_{-0.94}^{+0.89}$ &  WD     &         4.75  & $_{-0.58}^{+0.55}$ &  D      \\
$\log\,$$\Delta p_{\rm c}$    &    4.0    &       2.52  & $_{-1.68}^{+1.63}$ &  L      &         2.33  & $_{-1.68}^{+1.89}$ &  L      &         2.58  & $_{-1.81}^{+1.73}$ &  L       &         2.36  & $_{-1.68}^{+1.84}$ &  L      &         2.55  & $_{-1.84}^{+1.93}$ &  L      \\
$\log\,$$\tau_{\rm c}$        &    1.0    &      -0.07  & $_{-1.99}^{+1.95}$ &  NC     &         0.44  & $_{-2.13}^{+1.69}$ &  NC     &         0.32  & $_{-2.14}^{+1.65}$ &  NC      &         1.19  & $_{-0.34}^{+1.09}$ &  WD     &         1.05  & $_{-0.22}^{+0.49}$ &  C      \\
$\log\,$$f_{\rm c}$           &    -0.3   &      -1.32  & $_{-1.19}^{+1.04}$ &  NC     &        -1.38  & $_{-1.06}^{+0.93}$ &  NC     &        -1.17  & $_{-1.27}^{+0.88}$ &  NC      &        -0.27  & $_{-0.44}^{+0.22}$ &  WD     &        -0.32  & $_{-0.40}^{+0.24}$ &  WD     \\
$a$                           &    1.0    &       3.34  & $_{-2.37}^{+3.83}$ &  NC     &         2.95  & $_{-2.07}^{+3.53}$ &  WD     &         3.28  & $_{-2.52}^{+4.31}$ &  WD      &         2.74  & $_{-1.98}^{+3.94}$ &  WD     &         3.84  & $_{-2.78}^{+3.46}$ &  NC     \\
$\alpha$                      &    90.0   &      65.79  & $_{-42.59}^{+47.86}$ &  NC   &         57.5  & $_{-45.14}^{+53.47}$ &  L    &        62.08  & $_{-43.26}^{+48.52}$ &  L     &        68.37  & $_{-46.54}^{+59.77}$ &  L    &        67.34  & $_{-44.39}^{+42.43}$ &  L    \\
\bottomrule
\end{tabular}
\end{threeparttable}
}
\end{table}


\begin{table}
\centering
\resizebox{1.\textwidth}{!}{%
\begin{threeparttable}
\caption{Retrieval Results Comparison and Associated Level of Constraint for Observations Conducted in Different Spectral Coverages at S/N = 15}\label{tab:ret_results_SNR15}
\begin{tabular}{lS|S@{\hspace{-0.17cm}}l@{}c|S@{\hspace{-0.17cm}}l@{}c|S@{\hspace{-0.17cm}}l@{}c|S@{\hspace{-0.17cm}}l@{}c|S@{\hspace{-0.17cm}}l@{}c|}
\toprule
Parameter & Input & \multicolumn{3}{c}{Red} & \multicolumn{3}{c}{Blue \& Red} & \multicolumn{3}{c}{Visible} & \multicolumn{3}{c}{NIR} & \multicolumn{3}{c}{vNIR} \\
 \midrule
$\log\,$$f_{\rm N_2}$         &   -0.11   &      -5.14  & $_{-3.30}^{+3.53}$ &  NC     &        -3.61  & $_{-4.47}^{+2.53}$ &  NC     &        -4.74  & $_{-3.62}^{+3.07}$ &  NC     &      -4.96  & $_{-3.17}^{+3.37}$ &  NC       &       -4.81  & $_{-3.35}^{+3.22}$ &  NC     \\
$\log\,$$f_{\rm O_2}$         &   -0.68   &      -4.67  & $_{-3.38}^{+2.97}$ &  NC     &         -4.7  & $_{-3.38}^{+3.02}$ &  NC     &        -0.77  & $_{-0.57}^{+0.46}$ &  WD     &      -4.33  & $_{-3.94}^{+3.25}$ &  NC       &       -0.76  & $_{-0.52}^{+0.41}$ &  C      \\
$\log\,$$f_{\rm H_2O}$        &   -2.52   &      -1.68  & $_{-1.05}^{+0.94}$ &  L      &         -1.7  & $_{-0.91}^{+0.83}$ &  L      &        -2.58  & $_{-0.51}^{+0.46}$ &  C      &      -1.79  & $_{-1.05}^{+1.10}$ &  L        &       -2.55  & $_{-0.45}^{+0.47}$ &  C      \\
$\log\,$$f_{\rm CO_2}$        &    -3.4   &      -5.45  & $_{-3.17}^{+3.56}$ &  NC     &        -5.03  & $_{-3.46}^{+3.36}$ &  NC     &         -5.4  & $_{-3.42}^{+3.07}$ &  NC     &      -5.92  & $_{-2.81}^{+2.70}$ &  L        &       -5.83  & $_{-2.88}^{+2.38}$ &  L      \\
$\log\,$$f_{\rm CH_4}$        &    -5.7   &      -6.68  & $_{-2.36}^{+2.52}$ &  L      &        -6.81  & $_{-2.21}^{+2.64}$ &  L      &        -7.23  & $_{-1.85}^{+2.34}$ &  L      &      -7.49  & $_{-1.61}^{+2.02}$ &  L        &       -7.24  & $_{-1.82}^{+1.70}$ &  L      \\
$\log\,$$f_{\rm O_3}$         &   -6.15   &      -6.22  & $_{-2.51}^{+2.67}$ &  NC     &        -7.15  & $_{-1.99}^{+1.09}$ &  L      &        -6.17  & $_{-0.32}^{+0.32}$ &  C      &      -5.82  & $_{-2.83}^{+2.33}$ &  NC       &        -6.2  & $_{-0.30}^{+0.36}$ &  C      \\
$\log\,$$p_{\rm surf}$        &    5.0    &       4.75  & $_{-0.71}^{+0.77}$ &  WD     &         4.63  & $_{-0.77}^{+0.95}$ &  WD     &         5.02  & $_{-0.52}^{+0.55}$ &  D      &       5.12  & $_{-0.88}^{+1.26}$ &  WD       &        5.17  & $_{-0.46}^{+0.56}$ &  D      \\
$T$                           &   255.0   &     251.43  & $_{-44.11}^{+64.57}$ &  D    &       281.82  & $_{-74.27}^{+107.55}$ &  WD  &       269.03  & $_{-58.94}^{+75.16}$ &  WD   &     241.97  & $_{-35.48}^{+39.22}$ &  C      &      256.76  & $_{-34.37}^{+37.68}$ &  C    \\
$\log\,$$A_{\rm surf}$        &    -1.3   &      -0.69  & $_{-0.73}^{+0.47}$ &  NC     &        -1.07  & $_{-0.52}^{+0.45}$ &  L      &        -0.68  & $_{-0.50}^{+0.24}$ &  WD     &      -1.06  & $_{-0.65}^{+0.65}$ &  NC       &       -1.22  & $_{-0.46}^{+0.44}$ &  L      \\
$\log\,$$R_{\rm p}$           &    0.0    &       0.39  & $_{-0.50}^{+0.37}$ &  WD     &         0.36  & $_{-0.42}^{+0.36}$ &  WD     &         0.39  & $_{-0.36}^{+0.32}$ &  WD     &       0.33  & $_{-0.50}^{+0.39}$ &  WD       &        0.48  & $_{-0.45}^{+0.30}$ &  WD     \\
$\log\,$$M_{\rm p}$           &    0.0    &        0.9  & $_{-1.02}^{+0.78}$ &  NC     &         0.92  & $_{-0.94}^{+0.77}$ &  NC     &         0.78  & $_{-1.01}^{+0.83}$ &  NC     &       0.92  & $_{-0.98}^{+0.74}$ &  NC       &        0.92  & $_{-1.05}^{+0.71}$ &  NC     \\
$\log\,$$p_{\rm c}$           &    4.78   &       3.01  & $_{-2.11}^{+1.46}$ &  L      &         2.65  & $_{-1.82}^{+1.69}$ &  L      &         3.15  & $_{-2.05}^{+1.59}$ &  L      &       4.52  & $_{-0.72}^{+0.68}$ &  WD       &         4.8  & $_{-0.41}^{+0.41}$ &  C      \\
$\log\,$$\Delta p_{\rm c}$    &    4.0    &       2.48  & $_{-1.66}^{+1.61}$ &  L      &         2.25  & $_{-1.49}^{+1.63}$ &  L      &         2.58  & $_{-1.70}^{+1.75}$ &  L      &       2.45  & $_{-1.70}^{+1.69}$ &  L        &        2.65  & $_{-1.74}^{+1.63}$ &  L      \\
$\log\,$$\tau_{\rm c}$        &    1.0    &       0.58  & $_{-2.29}^{+1.52}$ &  NC     &        -0.04  & $_{-2.05}^{+1.71}$ &  NC     &         0.21  & $_{-2.09}^{+1.54}$ &  NC     &       1.08  & $_{-0.22}^{+0.96}$ &  WD       &        1.01  & $_{-0.16}^{+0.33}$ &  C      \\
$\log\,$$f_{\rm c}$           &    -0.3   &      -1.16  & $_{-1.27}^{+1.02}$ &  NC     &        -1.41  & $_{-1.05}^{+0.99}$ &  NC     &        -1.21  & $_{-1.17}^{+0.92}$ &  NC     &      -0.23  & $_{-0.38}^{+0.17}$ &  WD       &       -0.31  & $_{-0.36}^{+0.21}$ &  WD     \\
$a$                           &    1.0    &       3.02  & $_{-2.19}^{+3.79}$ &  NC     &         1.62  & $_{-1.04}^{+2.56}$ &  WD     &          2.6  & $_{-1.75}^{+3.47}$ &  WD     &       2.28  & $_{-1.61}^{+3.50}$ &  WD       &        3.68  & $_{-2.53}^{+3.71}$ &  NC     \\
$\alpha$                      &    90.0   &      54.88  & $_{-36.92}^{+60.46}$ &  L    &        60.01  & $_{-40.02}^{+57.13}$ &  L    &        66.97  & $_{-46.20}^{+49.47}$ &  L    &      77.03  & $_{-55.61}^{+54.85}$ &  L      &        63.0  & $_{-44.11}^{+42.09}$ &  L    \\
\bottomrule
\end{tabular}
\end{threeparttable}
}
\end{table}


\begin{table}
\centering
\resizebox{1.\textwidth}{!}{%
\begin{threeparttable}
\caption{Retrieval Results Comparison and Associated Level of Constraint for Observations Conducted in Different Spectral Coverages at S/N = 20}\label{tab:ret_results_SNR20}
\begin{tabular}{lS|S@{\hspace{-0.17cm}}l@{}c|S@{\hspace{-0.17cm}}l@{}c|S@{\hspace{-0.17cm}}l@{}c|S@{\hspace{-0.17cm}}l@{}c|S@{\hspace{-0.17cm}}l@{}c|}
\toprule
Parameter & Input & \multicolumn{3}{c}{Red} & \multicolumn{3}{c}{Blue \& Red} & \multicolumn{3}{c}{Visible} & \multicolumn{3}{c}{NIR} & \multicolumn{3}{c}{vNIR} \\
 \midrule
$\log\,$$f_{\rm N_2}$         &   -0.11   &      -4.43  & $_{-3.71}^{+2.85}$ &  NC     &        -4.93  & $_{-3.26}^{+3.49}$ &  NC     &         -4.7  & $_{-3.43}^{+3.06}$ &  NC      &       -5.01  & $_{-3.49}^{+3.20}$ &  NC      &        -5.1  & $_{-3.47}^{+3.28}$ &  NC     \\
$\log\,$$f_{\rm O_2}$         &   -0.68   &      -4.85  & $_{-3.71}^{+3.14}$ &  NC     &        -4.95  & $_{-3.30}^{+3.07}$ &  NC     &        -0.68  & $_{-0.42}^{+0.38}$ &  C       &       -3.24  & $_{-4.45}^{+2.62}$ &  NC      &       -0.77  & $_{-0.40}^{+0.35}$ &  C      \\
$\log\,$$f_{\rm H_2O}$        &   -2.52   &      -1.56  & $_{-0.97}^{+0.98}$ &  L      &        -1.67  & $_{-0.98}^{+0.88}$ &  L      &        -2.52  & $_{-0.45}^{+0.40}$ &  C       &       -1.82  & $_{-1.00}^{+1.06}$ &  L       &       -2.57  & $_{-0.41}^{+0.38}$ &  C      \\
$\log\,$$f_{\rm CO_2}$        &    -3.4   &      -5.38  & $_{-2.96}^{+3.30}$ &  NC     &        -5.15  & $_{-3.26}^{+3.11}$ &  NC     &        -5.04  & $_{-3.24}^{+3.15}$ &  NC      &       -5.72  & $_{-2.85}^{+2.44}$ &  L       &       -6.53  & $_{-2.38}^{+2.47}$ &  L      \\
$\log\,$$f_{\rm CH_4}$        &    -5.7   &      -6.71  & $_{-2.28}^{+2.20}$ &  L      &         -6.5  & $_{-2.32}^{+2.16}$ &  L      &         -6.8  & $_{-2.03}^{+2.28}$ &  L       &       -7.11  & $_{-1.91}^{+1.54}$ &  L       &       -7.44  & $_{-1.74}^{+1.70}$ &  L      \\
$\log\,$$f_{\rm O_3}$         &   -6.15   &      -6.11  & $_{-2.46}^{+2.88}$ &  NC     &        -6.78  & $_{-2.14}^{+0.89}$ &  L      &        -6.13  & $_{-0.28}^{+0.31}$ &  C       &       -6.29  & $_{-2.51}^{+2.80}$ &  NC      &       -6.22  & $_{-0.25}^{+0.30}$ &  C      \\
$\log\,$$p_{\rm surf}$        &    5.0    &       4.72  & $_{-0.65}^{+0.66}$ &  NC     &          4.6  & $_{-0.76}^{+0.83}$ &  NC     &         5.01  & $_{-0.42}^{+0.47}$ &  C       &        5.08  & $_{-0.79}^{+1.36}$ &  NC      &        5.16  & $_{-0.41}^{+0.55}$ &  C      \\
$T$                           &   255.0   &     255.83  & $_{-39.19}^{+41.90}$ &  C    &       257.27  & $_{-52.79}^{+69.80}$ &  WD   &       263.67  & $_{-40.87}^{+53.22}$ &  WD    &      248.07  & $_{-31.07}^{+29.44}$ &  C     &      254.56  & $_{-27.08}^{+31.44}$ &  C    \\
$\log\,$$A_{\rm surf}$        &    -1.3   &      -0.67  & $_{-0.88}^{+0.48}$ &  NC     &        -1.17  & $_{-0.50}^{+0.49}$ &  L      &        -0.76  & $_{-0.63}^{+0.32}$ &  WD      &       -1.27  & $_{-0.53}^{+0.73}$ &  NC      &       -1.28  & $_{-0.46}^{+0.38}$ &  L      \\
$\log\,$$R_{\rm p}$           &    0.0    &       0.28  & $_{-0.41}^{+0.43}$ &  WD     &          0.4  & $_{-0.35}^{+0.31}$ &  WD     &         0.33  & $_{-0.35}^{+0.30}$ &  WD      &        0.38  & $_{-0.51}^{+0.34}$ &  WD      &        0.47  & $_{-0.47}^{+0.28}$ &  WD     \\
$\log\,$$M_{\rm p}$           &    0.0    &       0.95  & $_{-1.10}^{+0.75}$ &  NC     &         0.97  & $_{-1.07}^{+0.75}$ &  NC     &         0.66  & $_{-0.90}^{+0.83}$ &  NC      &        1.02  & $_{-1.04}^{+0.67}$ &  NC      &         0.8  & $_{-1.00}^{+0.75}$ &  NC     \\
$\log\,$$p_{\rm c}$           &    4.78   &       3.37  & $_{-2.21}^{+1.16}$ &  L      &         3.13  & $_{-2.00}^{+1.55}$ &  L      &          4.3  & $_{-1.80}^{+0.65}$ &  WD      &        4.58  & $_{-0.61}^{+0.61}$ &  D       &        4.81  & $_{-0.33}^{+0.34}$ &  C      \\
$\log\,$$\Delta p_{\rm c}$    &    4.0    &       2.13  & $_{-1.42}^{+1.60}$ &  L      &         1.99  & $_{-1.28}^{+1.73}$ &  L      &         2.27  & $_{-1.62}^{+1.79}$ &  L       &        2.09  & $_{-1.52}^{+1.73}$ &  L       &        2.33  & $_{-1.56}^{+1.74}$ &  L      \\
$\log\,$$\tau_{\rm c}$        &    1.0    &       0.53  & $_{-2.30}^{+1.48}$ &  NC     &         0.36  & $_{-2.00}^{+1.49}$ &  NC     &         0.67  & $_{-2.01}^{+1.27}$ &  NC      &        1.03  & $_{-0.16}^{+0.59}$ &  C       &         1.0  & $_{-0.12}^{+0.23}$ &  C      \\
$\log\,$$f_{\rm c}$           &    -0.3   &      -1.12  & $_{-1.31}^{+0.99}$ &  NC     &        -1.26  & $_{-1.14}^{+1.00}$ &  NC     &        -0.95  & $_{-1.34}^{+0.72}$ &  NC      &       -0.23  & $_{-0.41}^{+0.18}$ &  WD      &       -0.26  & $_{-0.29}^{+0.19}$ &  C      \\
$a$                           &    1.0    &       2.06  & $_{-1.41}^{+3.40}$ &  WD     &         1.73  & $_{-1.03}^{+1.84}$ &  WD     &         2.24  & $_{-1.47}^{+2.45}$ &  WD      &        2.31  & $_{-1.58}^{+3.75}$ &  WD      &        3.68  & $_{-2.47}^{+3.63}$ &  NC     \\
$\alpha$                      &    90.0   &      73.43  & $_{-49.44}^{+54.62}$ &  L    &        62.75  & $_{-46.09}^{+57.70}$ &  L    &        70.88  & $_{-49.12}^{+52.92}$ &  L     &       75.26  & $_{-48.07}^{+49.64}$ &  L     &       66.28  & $_{-45.75}^{+37.97}$ &  L    \\
\bottomrule
\end{tabular}
\end{threeparttable}
}
\end{table}


\bibliography{biblio}{}
\bibliographystyle{aasjournal}
\listofchanges

\end{document}